\documentclass[twocolumn,aps,prl,preprintnumbers,showpacs,nofootinbib]{revtex4}
\usepackage[utf8]{inputenc}
\usepackage[english]{babel}
\usepackage{amsmath}
\usepackage{amsfonts}
\usepackage{amssymb}
\usepackage{epsfig}
\usepackage{graphics,psfrag,rotating}
\usepackage{graphicx}
\usepackage{dcolumn}
\usepackage{bm}
\usepackage{epstopdf}
\usepackage{color}
\usepackage[usenames,dvipsnames,svgnames]{xcolor}
\usepackage[colorlinks=true,
            linkcolor=red,
            urlcolor=gray,
            citecolor=blue]{hyperref}
  \usepackage{hyperref}

\newcommand{\lsim}   {\mathrel{\mathop{\kern 0pt \rlap
{\raise.2ex\hbox{$<$}}}
 \lower.9ex\hbox{\kern-.190em $\sim$}}}
\newcommand{\gsim}   {\mathrel{\mathop{\kern 0pt \rlap
{\raise.2ex\hbox{$>$}}}
\lower.9ex\hbox{\kern-.190em $\sim$}}}

\def\3nab{\tilde{\nabla}}

\def\hsp5{\hspace{5mm}}

\def\case#1/#2{\textstyle\frac{#1}{#2}}

\def\ber {\begin{eqnarray}}
\def\eer {\end{eqnarray}}
\def\bea {\begin{eqnarray}}
\def\eea {\end{eqnarray}}

\def\bc {\begin{center}}
\def\ec {\end{center}}
\def\case#1/#2{\frac{#1}{#2}}

\newcommand{\bw}{\begin{widetext}}
\newcommand{\ew}{\end{widetext}}

\newcommand{\be}{\begin{equation}}
\newcommand{\bse}{\begin{subequation}}
\newcommand{\ese}{\end{subequation}}
\newcommand{\ee}{\end{equation}}
\newcommand{\eei}{\end{eqnarray}\indent\indent}
\newcommand{\ba}{\begin{array}}
\newcommand{\ea}{\end{array}}
\newcommand{\bal}{\begin{eqnarray}}
\newcommand{\eal}{\end{eqnarray}}

\def\case#1/#2{\textstyle\frac{#1}{#2} }

\newcommand{\bver}{\begin{verbatim}}
\newcommand{\ever}{\end{verbatim}}

\begin{document}

%\preprint{APS/123-QED}

\title{The interaction of  neutrinos with  Phantom, Quintessence, and Quintum scalar fields and its effect on the formation of structures in the early Universe.}
\author{
 Muhammad Yarahmadi\footnote{Email: Yarahmadimohammad10@gmail.com}
Amin Salehi\footnote{Email: Salehi.a@lu.ac.ir}
}
\affiliation{Department of Physics, Lorestan University, Khoramabad, Iran}

\date{\today}% It is always \today, today,

\begin{abstract}
Despite the fact that the mass of the neutrinos is so small, they are produced in such vast numbers in the early Universe that their mass induces subtle effects on cosmological observables, primarily the growth of structure and the expansion history in the Universe.
We consider the models where neutrino interacts with dark energy scalar field models; phantom, quintessence, and quintom. Also, we obtained the $z_{\rm nr}$ (the redshift at which a  mass of neutrino  $m_{\nu}$ will become non-relativistic) and surveyed the effect of non-relativistic neutrinos on the structure formation.  The data used in this paper are Pantheon + Analysis catalog, CMB, and BAO data. We obtained coupling constant $\beta$ for neutrino and three scalar fields and found that larger $\beta$ values will generally lead to larger neutrino mass in the  Universe. For combination data, we found that the total mass of neutrino  $\sum m_{\nu}< 0.1197$eV $(95\% $ Confidence Level (C.L.)  for quintom model and $\sum m_{\nu}< 0.121 $eV $(95\% $ Confidence Level (C.L.) for phantom model and $\sum m_{\nu}< 0.122$eV $(95\% $ Confidence Level (C.L.) for quintessence model. These results are in broad agreement with the results of Planck 2018 where the total neutrino mass is $\sum m_{\nu}<0.12$eV ($95\%$ C.L., TT, TE, EE+lowE+lensing+BAO). Using the neutrino mass obtained from different models, we calculated  $z_{\rm nr}$ and co-moving wave number $k_{\rm nr}$ and showed that neutrinos played a role on the structure formation in the early Universe.
\end{abstract}


%04.50.Kd	Modified theories of gravity
%04.25.Nx 	Post-Newtonian approximation; perturbation theory; related approximations
% 95.36.+x 	Dark energy (see also 98.80.-k Cosmology)
% 98.80.-k	 Cosmology (see also section 04 General relativity and gravitation; for origin and evolution of galaxies, see 98.62.Ai; for elementary particle and nuclear processes, see 95.30.Cq; for dark matter, see 95.35.+d; for dark energy, see 95.36.+x; for superclusters and large-scale structure of the Universe, see 98.65.Dx)
%
%
% 04.50.-h 	Higher-dimensional gravity and other theories of gravity (see also 11.25.Mj Compactification and four-dimensional models, 11.25.Uv D branes)
%04.20.-q	         Classical general relativity (see also 02.40.-k Geometry, differential geometry, and topology)
%04.20.Cv	Fundamental problems and general formalism
%04.50.-h	         Higher-dimensional gravity and other theories of gravity (see also 11.25.Mj Compactification and four-dimensional models, 11.25.Uv D branes)
% 98.80.Jk        Mathematical and relativistic aspects of cosmology
% 04.20.Jb        Exact solutions


\maketitle

\section{introduction}
The model space of beyond $\Lambda$CDM cosmologies includes a range of scenarios
featuring interactions between dark energy and other species. (\cite{Wetterich}; \cite{Amendola}; \cite{Farrar}; \cite{Pettorino}; \cite{Baldi}), Models can be constructed in which dark energy interacts with matter-energy fields. These are generally confined to couplings to cold dark matter or neutrinos. There are a number of interesting models for neutrinos acquiring a growing mass by coupling to a scalar field. These models have a scalar fields $\phi $  playing
the role of dark energy coupled to the neutrinos in such a way that they provide a 'trigger' that causes the scalar field to leave a scaling regime and enter an inflationary regime (\cite{Ade};\cite{Riess};\cite{Perlmutter};\cite{Alam}). If we want to explain this issue  by general relativity in four dimensions, we  have two ways to describe this recent cosmic expansion: one way is to modify the gravitational part of Einstein's equations (\cite{Sotiriou};\cite{Nojiri}), and the other being  to modify the  universe's contents. The simplest candidate for dark energy is the cosmological constant with equation of state parameter $\omega = -1$, which fits well with the observational evidence. If the cosmological constant is the reason for the recent cosmic acceleration, we must find a mechanism that can obtain a very small amount of it which is consistent with the observational evidence. Unfortunately, the cosmological constant has several serious problems, such as  fine-tuning problem (\cite{Carroll};\cite{Sahni};\cite{Padmanabhan}). There are several  candidates for dark energy such as scalar fields. A canonical scalar field is called the quintessence \cite{Ratra} and non - canonical scalar field  is a phantom field with negative kinetic energy (\cite{Copeland}, tachyon field is derived from string theory \cite{Sen}, a scalar field with generalized kinetic energy called K-Essence field  (\cite{Armendariz};\cite{Chiba}) and Chaplygn gas\cite{Bilic}. The quintessence and phantom fields in a combined model are yet another candidate for dark energy,  called the quintom model(\cite{Guo};\cite{Nozari};\cite{Rashidi}). As we know, in the quintessence model of dark energy, the  equation of state parameter always remains greater than -1. Moreover, in the phantom model, the  equation of state parameter is always less than -1. An important feature of the Quintom model is that in this model the equation of state parameter can cross the phantom boundary. Because of this, the Quintom model seems to be an interesting candidate for dark energy. While neutrinos play an important role in the early Universe cosmology, their impact on the late universe is relatively minor in $\Lambda$CDM. There are some cosmological models, however, in which neutrinos are given a central role in the late universe by means of a coupling to the dark energy.
Besides, the Standard Model of Particle Physics does not explain  how the neutrino masses are generated. Cosmology has helped us to come  closer to explaining neutrino masses. In particular, cosmological observations are sensitive to the imprint of neutrinos on structure formation, because neutrinos suppress structure formation on small scales and slow down the growth of structure on all scales. Despite the fact that
the mass of the neutrinos is so small, they are produced in such vast numbers in the early Universe that their mass induces subtle effects on cosmological observables, primarily the growth of structure and the expansion history in the Universe. Throughout the history of the Universe, neutrinos from the early Universe have evolved from a relativistic phase at very early times to a massive-particle behavior at later times \cite{Julien}. Initially,  neutrinos kinetic energy dominates over their rest-mass energy, and as a consequence, neutrinos can be described as massless particles fully characterized by their temperature. As the Universe cools down, the kinetic energy decreases and neutrinos undergo a transition to a non-relativistic phase with a non-negligible mass. We are interested in highlighting the role of neutrinos. In this work, we implement a cosmological model proposed by observations to put a constraint on the total neutrino mass(\cite{Caldwell};\cite{Zhao}) and the effect of neutrinos on structure formation.
Scalar field plays the role of dark energy and is responsible for the accelerating expansion of the Universe.

\section{The Models}
\subsection{Phantom model}
In this section, we study all three scalar fields separately. We start with the phantom dark energy.
Evidence from observational data suggests that the universe is currently in a narrow band near parameter $\omega=-1$ and may be well below this value (which lies in the so-called phantom regime), an area called the Phantom Age.
Phantom model is the non-canonical scalar field which is very similar to quintessence model (canonical scalar fields) except for the fact that it has negative kinetic energy.  The Lagrangian of this model is:
\begin{equation}\label{fried}
	\begin{split}
		L_{\sigma}=\dfrac{1}{2}{\partial _\mu }\sigma {\partial ^\mu }\sigma -V_{\sigma}
	\end{split}
\end{equation}
where $V(\sigma)=V_{0}\exp^{-\lambda\kappa\sigma}$ is a self-interacting potential in which $\rm V_{0}$ is the potential at present, $\lambda$ denotes a dimensionless parameter
that determines the slope of the potential, and  $\kappa  = \sqrt {\frac{{8\pi G}}{{{c^4}}}} $.  Assuming  a flat space in  FLRW  metric  filled with baryons, radiation, dark matter, dark energy and neutrinos. We start with the Friedmann equations as:
\begin{equation}\label{fried}
	\begin{split}
		3H^{2}=\kappa^{2}\left(\rho_{\rm b}+\rho_{\rm c}+\rho_{\rm r}+\rho_{\nu}-\frac{1}{2}\dot{\sigma}^{2}+V(\sigma)\right)
	\end{split}
\end{equation}
where $\rho_{\rm b}$ is the baryon density, $\rho_{\rm c}$ is the cold dark matter density, $\rho_{\rm r}$ is radiation density and $\rho_{\nu}$ is density of neutrino.
\begin{equation}\label{fried}
	\begin{split}
		2\dot{H}+3H^{2}=\kappa^{2}(-\omega_{\rm b}\rho_{\rm b}-\omega_{\rm c}\rho_{\rm c}+\omega_{\rm r}\rho_{\rm r}\\-\omega_{\nu}\rho_{\nu}+\frac{1}{2}\dot{\sigma}^{2}+
		V(\sigma))
	\end{split}
\end{equation}
The energy density conservation equations are:
\begin{equation}\label{fried}
	\begin{split}
		\dot{\rho}_{\sigma}+3H\rho_{\sigma}(1+\omega_{\sigma})=-\beta\rho_{\nu}(1-3\omega_{\nu})\dot{\sigma}
	\end{split}
\end{equation}
The  action of neutrino-scalar field interaction resulting from \cite{Bean} can also be considered in the context of the coupled scalar field model(\cite{Wetterich1};\cite{Amendola};\cite{Wetterich})
By employing the Fermi-Dirac distribution for neutrinos whose masses
$m_{\nu}(\sigma)$ are
$\sigma$  dependent and  also in thermal equilibrium with temperature $T(\nu)$,  one obtains

\begin{equation}
	\begin{split}
		{\rho _\nu } = \frac{{{T^4}_{(\nu )}}}{{{\pi ^2}}}\int_0^\infty  {\frac{{dx{x^2}\sqrt {{x^2} + {\xi ^2}} }}{{{e^x} + 1}}}
	\end{split}
\end{equation}\\

\begin{equation}
	\begin{split}
		{p_\nu } = \frac{{{T^4}_{(\nu )}}}{{3{\pi ^2}}}\int_0^\infty  {\frac{{dx{x^4}}}{{{e^x} + 1\sqrt {{x^2} + {\xi ^2}} }}}
	\end{split}
\end{equation}\\
where $
\xi  = \frac{{{m_\nu }(\sigma )}}{{{T_{(\nu )}}}}$. By using above equation one finds

\begin{equation}
	\begin{split}
		\dot{\rho_{\nu}}+3H(\rho_{\nu}+p_{\nu})=\beta\dot{\sigma}(\rho_{\nu}-3p_{\nu})
	\end{split}
\end{equation}
Also, \cite{Brookfield} used the same way to drive eq(7).

\begin{equation}\label{fried}
	\begin{split}
		\dot{\rho}_{\nu}+3H\rho_{\nu}(1+\omega_{\nu})=\beta\rho_{\nu}(1-3\omega_{\nu})\dot{\sigma}
	\end{split}
\end{equation}
Note that if $\omega_{\nu} = \frac{1}{3}$, the coupling
vanishes making neutrinos non-interacting particles.

\begin{equation}\label{fried}
	\begin{split}
		\dot{\rho}_{\rm c}+3H\rho_{\rm c}=-\alpha\rho_{\rm c}\dot{\sigma}
	\end{split}
\end{equation}

\begin{equation}\label{fried}
	\begin{split}
		\dot{\rho}_{\rm b}+3H\rho_{\rm b}=0
	\end{split}
\end{equation}

\begin{equation}\label{fried}
	\begin{split}
		\dot{\rho}_{\rm r}+4H\rho_{\rm r}=0
	\end{split}
\end{equation}
respectively. Where the coupling parameter $\beta$ can, in general, be some function of $\sigma$.
The coupling  plays a role only when the neutrinos are non-relativistic, since relativistic
neutrinos have $P_{\upsilon} \approx \frac{\rho_{\upsilon}}{3}$; therefore,  the right-hand sides of Eqs. (3.4) and (3.7) are both
negligible such that the standard, uncoupled conservation equations are recovered. Since  dark energy is modeled as a scalar field $\sigma$ its energy density and pressure are given by
\begin{equation}
	\begin{split}
		\rho_{\sigma}=\frac{-1}{2}\dot{\sigma}^{2}+V(\sigma), \ \
		p_{\sigma}=\frac{-1}{2}\dot{\sigma}^{2}-V(\sigma)
	\end{split}
\end{equation}
where $V(\sigma)$ denotes the potential of the scalar field. In this paper, we consider the coupling between  dark energy and neutrino as the following evolution equation.
In addition, from the resulting 
equations of the scalar field :
\begin{equation}\label{fried}
	\begin{split}
		\ddot{\sigma}=-\lambda V(\sigma)-\frac{3}{2}H\dot{\sigma}(1+\omega_{\sigma})-\frac{3HV}{\dot{\sigma}}(1+\omega_{\phi})\\
		+\beta\rho_{\nu}(1-3\omega_{\nu})
	\end{split}
\end{equation}
where as before $\dot{a}$  means differentiation with respect to the coordinate time t. The EoS of
the scalar field is now given by:
\begin{equation}\label{fried}
	\begin{split}	
		\omega_{\sigma}=\frac{\frac{1}{2}\dot{\sigma}^{2}+V(\sigma)}{\frac{1}{2}\dot{\sigma}^{2}-V(\sigma)}
	\end{split}
\end{equation}
The above equations are a nonlinear set of second-order differential equations that can only be solved analytically for certain cases. To simplify the equations,  we can introduce a number of new variables  to turn the second-order differential equations into a set of first-order equations. There are several reasons for this, including:\\
1- Systems of the first order for the numerical solution are very simple and convenient. On the other hand, it allows us to study the behavior of the system in phase space.
2- In the numerical solution of first-order equations, unlike high-order equations, which require more than one condition for each equation, only one initial condition is required.\\
3- Most importantly, the first-order dynamics can be described on the phase space and it is possible to check the stability of the system.

If we considered the standard model, the right side of the equation would be zero, and the neutrinos  effectively uncoupled. We consider an exponential potential $V(\sigma)=V_0\;e^{-\lambda k\sigma}$, where $\lambda$ is a dimensionless parameter
that determines the slope of the potential. The motivation for choosing these functions have been investigated in \cite{Wetterich}.
Furthermore, we define $ \omega = \frac{P_{\nu}}{\rho_{\nu}}$. In order to simplify the field equations, we introduce the following new variables,
\begin{equation}\label{fried}
	\begin{split}
		\xi_{1}=\frac{\kappa^{2}\rho_{\rm b}}{3H^{2}} \ \ \ \ \xi_{2}=\frac{\kappa^{2}\rho_{\nu}}{3H^{2}} \ \ \ \ \xi_{3}=\frac{\kappa^{2}\rho_{\rm r}}{3H^{2}} \ \ \ \xi_{4}=\frac{\kappa^{2}\rho_{c}}{3H^{2}} \ \ \ \ \\ \xi_{5}=-\frac{\kappa\dot{\sigma}}{\sqrt{6}H} \ \ \
		\xi_{6}=\frac{\kappa^{2}V(\sigma)}{3H^{2}}
	\end{split}
\end{equation}
In term of new variable the Friedmann equations (2) puts a constraint on new variables as
\begin{equation}\label{cons}
	\begin{split}
		\xi_{6}=1-\xi_{1}-\xi_{2}-\xi_{3}-\xi_{4}+\xi_{5}^{2}
	\end{split}
\end{equation}
Therefore, the equations are simplified as follows:
\begin{equation}\label{fried}
	\frac{d\xi_{1}}{dN}=-3\xi_{1}-2\xi_{1}\frac{\dot{H}}{H^{2}} \ \ \ \ \ \
\end{equation}
\begin{equation}\label{fried}
	\frac{d\xi_{2}}{dN}=-\xi_{2}\left(3(1+\omega_\nu)+\sqrt{6}\xi_{5}\beta(1-3\omega_\nu)\right)-2\xi_{2}\frac{\dot{H}}{H^{2}} \ \ \
\end{equation}
\begin{equation}\label{fried}
	\frac{d\xi_{3}}{dN}=-4\xi_{3}-2\xi_{3}\frac{\dot{H}}{H^{2}} \ \ \
\end{equation}
\begin{equation}\label{fried}
	\frac{d\xi_{4}}{dN}=-\xi_4\left(3-\sqrt{6}\xi_5\alpha+2\frac{\dot{H}}{H^{2}}\right)
\end{equation}
\begin{equation}\label{fried}
	\frac{d\xi_{5}}{dN}=\frac{{3\lambda {\xi _6}}}{{\sqrt 6 }} - \frac{{3\beta {\xi _2}(1 - 3{\omega _\nu })}}{{\sqrt 6 }} + \frac{{9\alpha }}{{\sqrt 6 }}{\xi _4} + 3{\xi _5} - {\xi _5}\frac{{\dot H}}{{{H^2}}}
\end{equation}

Where, $N=\ln a$. In term of the new dynamical variable, we also have,
\begin{equation}\label{fried}
	\begin{split} \frac{\dot{H}}{H^{2}}=\frac{1}{2}\left(-3-\xi_3-3\omega_{\nu}\xi_{2}+3\xi_{5}^{2}+3\xi_{6}\right)
	\end{split}
\end{equation}
The above parameter is very important, since essential cosmological parameters such as deceleration parameters $q$ and effective equation of state (EoS) $w_{\rm eff}$ can be expressed in terms of this parameter as $q=-1-\frac{\dot{H}}{H^{2}}$ and $w_{\rm eff}=-1-\frac{2}{3}\frac{\dot{H}}{H^{2}}$. It is also used in calculating the luminosity distance. The equation related to the luminosity distance  is coupled by these parameters with the equations of the system. We used the same approach with\cite{Salehi}.\\

When neutrinos are non-relativistic, one needs to limit the value of $N_{\rm eff}$ accordingly. Also, the matter density must contain the neutrino contribution.
\begin{equation}
	{\Omega _{\rm m}} = {\Omega _{\nu}} + {\Omega _{\rm b}} + {\Omega _{\rm c}}
\end{equation}
where $\Omega_{\nu}$ is related to the sum of neutrino masses as
\begin{equation}
	{\Omega _\nu } = \frac{{{{\sum m }_\nu }}}{{94{h^2}}}
\end{equation}
In addition to photons, neutrinos and other relativistic degrees of freedom make up the total relativistic energy density in the early universe. $N_{\rm eff}$ parameter is used to show the effective number of relativistic species. Its standard value of 3.046 corresponds to the case of three generations of $N_{\rm eff}$ neutrinos with no additional dark radiation. Therefore, the total radiation energy density in the Universe is calculated  by

\begin{equation}
	\begin{split}
		\rho_{\rm r}=\rho_{\gamma}\left(1+\frac{7}{8}(\frac{4}{11})^{\frac{4}{3}}N_{\rm eff}\right)
	\end{split}
\end{equation}
where $\rho_{\gamma}$
is the energy density of photons.
This value of $N_{\rm eff}$  indicates that if it is higher than 3.046, there is a dark radiation that is other than three generations of neutrinos.

The expected linear correlation
between $\Omega_{\rm m}h^{2}$ and $N_{\rm eff}$ given by

\begin{equation}
	\begin{split}
		N_{\rm eff}=3.04+7.44\left(\frac{\Omega_{\rm m}h^{2}}{0.1308}\frac{3139}{1+z_{\rm eq}}-1\right)
	\end{split}
\end{equation}

\subsection{Quintessence model}
Quintessence is described as a canonical scalar field that is minimally coupled to gravity. Compared to other models of scalar fields such as phantoms, the simplest scenario is the scalar field. A variable slow field along with a potential can accelerate the universe. This mechanism is similar to slow inflation in the early universe; the difference is that non-relative matter (dark matter and baryon) cannot be ignored to properly discuss dark energy dynamics. The action which will then
represent our physical system is :

\begin{equation}\label{fried}
	\begin{split}
		S = \int {{d^4}} x\sqrt { - g} (\frac{R}{{2{\kappa ^2}}} + {L_m} + {L_\phi })	
	\end{split}
\end{equation}

where $ {L_\phi } $ is the canonical Lagrangian of a scalar ﬁeld $ \phi  $ uniquely given by
\begin{equation}\label{fried}
	\begin{split}
		{L_\phi } =  - \frac{1}{2}{\partial _\mu }\phi {\partial ^\mu }\phi  - V(\phi )	
	\end{split}
\end{equation}

Now, we analyze the quintessence model. We start with Friedmann and acceleration equation:(\cite{Ratra}; \cite{Wetterich}; \cite{Zlatev}; \cite{Peebles})

\begin{equation}\label{fried}
	\begin{split}
		3H^{2}=\kappa^{2}\left(\rho_{\rm b}+\rho_{\rm c}+\rho_{\rm r}+\rho_{\nu}+\frac{1}{2}\dot{\phi}^{2}+V(\phi)\right)	
	\end{split}
\end{equation}

\begin{equation}\label{fried}
	\begin{split}
		2\dot{H}+3H^{2}=-\kappa^{2}(\omega_{\rm b}\rho_{\rm b}+\omega_{\rm c}\rho_{\rm c}+\omega_{\rm r}\rho_{\rm r}+\omega_{\nu}\rho_{\nu}\\+\frac{1}{2}\dot{\phi}^{2}-V(\phi))
	\end{split}
\end{equation}
The evolution equations for
their energy densities are :
\begin{equation}\label{fried}
	\begin{split}
		\dot{\rho}_{\phi}+3H\rho_{\phi}(1+\omega_{\phi})=-\beta\rho_{\nu}(1-3\omega_{\nu})\dot{\phi}
	\end{split}
\end{equation}
\begin{equation}\label{fried}
	\begin{split}
		\dot{\rho}_{\nu}+3H\rho_{\nu}(1+\omega_{\nu})=\beta\rho_{\nu}(1-3\omega_{\nu})\dot{\phi}
	\end{split}
\end{equation}
while the Klein-Gordon equation  is:
\begin{equation}\label{fried}
	\begin{split}
		\ddot{\phi}=\lambda V(\phi)-\frac{3}{2}H\dot{\phi}(1+\omega_{\phi})-\frac{3HV}{\dot{\phi}}(1+\omega_{\phi})
		-\beta\rho_{\nu}\\(1-3\omega_{\nu})
	\end{split}
\end{equation}

We can define the energy density and pressure of the scalar field as follows:
\begin{equation}\label{fried}
	\begin{split}
		\rho_{\phi}=\frac{1}{2}\dot{\phi}^{2}+V(\phi)
	\end{split}
\end{equation}

\begin{equation}\label{fried}
	\begin{split}
		p_{\phi}=\frac{1}{2}\dot{\phi}^{2}-V(\phi)
	\end{split}
\end{equation}
The resulting equation of state is as follows:
\begin{equation}\label{fried}
	\begin{split}	
		\omega_{\phi}=\frac{\frac{1}{2}\dot{\phi}^{2}-V(\phi)}{\frac{1}{2}\dot{\phi}^{2}+V(\phi)}
	\end{split}
\end{equation}
where $\omega_{\phi}$ is a dynamically evolving parameter which can take values in the range $[-1 , 1]$.
We rewrite the cosmological equations (3.24), (3.25) and (3.26) into an autonomous system of
equations.
\begin{equation}\label{fried}
	\begin{split}
		\chi_{1}=\frac{\kappa^{2}\rho_{\rm b}}{3H^{2}} \ \ \ \ \chi_{2}=\frac{\kappa^{2}\rho_{\nu}}{3H^{2}} \ \ \ \ \chi_{3}=\frac{\kappa^{2}\rho_{\rm r}}{3H^{2}} \ \ \ \chi_{4}=\frac{\kappa^{2}\rho_{c}}{3H^{2}} \ \ \ \ \\ \chi_{5}=-\frac{\kappa\dot{\sigma}}{\sqrt{6}H} \ \ \
		\chi_{6}=\frac{\kappa^{2}V(\phi)}{3H^{2}}
	\end{split}
\end{equation}
we can derive the following dynamical system:
\begin{equation}\label{cons}
	\begin{split}
		\chi_{6}=1-\chi_{1}-\chi_{2}-\chi_{3}-\chi_{4}+\chi_{5}^{2}
	\end{split}
\end{equation}
Therefore, the equations are simplified as follows:
\begin{equation}\label{fried}
	\frac{d\chi_{1}}{dN}=-3\chi_{1}-2\chi_{1}\frac{\dot{H}}{H^{2}} \ \ \ \ \ \
\end{equation}
\begin{equation}\label{fried}
	\frac{d\chi_{2}}{dN}=-\chi_{2}\left(3(1+\omega_\nu)-\sqrt{6}\chi_{5}\beta(1-3\omega_\nu)\right)-2\chi_{2}\frac{\dot{H}}{H^{2}} \ \ \
\end{equation}
\begin{equation}\label{fried}
	\frac{d\chi_{3}}{dN}=-4\chi_{3}-2\chi_{3}\frac{\dot{H}}{H^{2}} \ \ \
\end{equation}
\begin{equation}\label{fried}
	\frac{d\chi_{4}}{dN}=-\chi_4\left(3+\sqrt{6}\chi_5\alpha+2\frac{\dot{H}}{H^{2}}\right)
\end{equation}
\begin{equation}\label{fried}
	\frac{d\chi_{5}}{dN}=\frac{{3\lambda {\chi_6}}}{{\sqrt 6 }} - \frac{{3\beta {\chi_2}(1 - 3{\omega _\nu })}}{{\sqrt 6 }} + \frac{{9\alpha }}{{\sqrt 6 }}{\chi_4} - 3{\chi_5} - {\chi_5}\frac{{\dot H}}{{{H^2}}}
\end{equation}

Where, $N=\ln a$. In terms of the new dynamical variables, we also have,
\begin{equation}\label{fried}
	\begin{split} \frac{\dot{H}}{H^{2}}=\frac{1}{2}\left(-3-\chi_3-3\omega_{\nu}\chi_{2}-3\chi_{5}^{2}+3\chi_{6}\right)
	\end{split}
\end{equation}
Using the relationships mentioned to estimate the mass of neutrinos and $N_{\rm eff}$ in the previous section, we have:

\subsection{Quintom model}
In the previous section we found that the equation of state  in the quintessence model must satisfy $\omega_{\rm de} \geq-1$, we also saw that in the phantom scalar field, EoS is limited to $\omega_{\rm de} <-1$. There seems to be no way to cross the phantom barrier (i.e. the cosmological constant $\omega_{\rm de} =-1$) using a scalar field. To cross this barrier, the Quintom model is used, which allows such a passage. This dark energy scenario causes the EoS to be larger than -1  and less than -1, which satisfies current observations. The simplest model is represented by a quantum Lagrangian consisting of two scalar fields, a canonical field $\phi$ (quintessence) and a phantom field $\sigma$:
\begin{equation}\label{fried}
	\begin{split}
		{L_{qu{\mathop{\rm int}} om}} =  - \frac{1}{2}{\partial _\mu }\phi {\partial ^\mu }\phi  + \frac{1}{2}{\partial _\mu }\sigma {\partial ^\mu }\sigma  - V(\sigma ,\phi )
	\end{split}
\end{equation}
where $V(\phi,\sigma)$ is
\begin{equation}\label{fried}
	\begin{split}
		V(\sigma,\phi)=V_{0}\exp^{-\lambda_{\phi}\kappa\phi-\lambda_{\sigma}\kappa\sigma}
	\end{split}
\end{equation}
a general potential for both  scalar fields and $ \phi $ and $ \sigma $ represents the quintessence and phantom fields, respectively. In above equation $\lambda_{\phi}$ and $\lambda_{\sigma}$ are constant values. The kinetic energy sign is positive for the quintessence model and negative for the phantom model. The cosmological equations are given by the Friedmann and acceleration equations: (\cite{Guo}; \cite{Zhang})
\begin{equation}\label{fried}
	\begin{split}
		3H^{2}=\kappa^{2}\left(\rho_{\rm b}+\rho_{\rm c}+\rho_{\rm r}+\rho_{\nu}+\frac{1}{2}\dot{\phi}^{2}+V(\sigma,\phi)-\frac{1}{2}\dot{\sigma}^{2}\right)
	\end{split}
\end{equation}

\begin{equation}\label{fried}
	\begin{split}
		2\dot{H}+3H^{2}=-\kappa^{2}(\omega_{\rm b}\rho_{\rm b}+\omega_{\rm c}\rho_{\rm c}+\omega_{\rm r}\rho_{\rm r}+\omega_{\nu}\rho_{\nu}+\\ \frac{1}{2}\dot{\phi}^{2}+V(\sigma,\phi)-\frac{1}{2}\dot{\sigma}^{2})
	\end{split}
\end{equation}
The energy density conservation equations are:
\begin{equation}\label{fried}
	\begin{split}
		\dot{\rho}_({\sigma,\phi})+3H\rho_{({\sigma,\phi})}(1+\omega_{({\sigma,\phi})})=-\beta\rho_{\nu}(1-3\omega_{\nu})(\dot{\sigma}+\dot{\phi})
	\end{split}
\end{equation}

\begin{equation}\label{fried}
	\begin{split}
		\dot{\rho}_{\nu}+3H\rho_{\nu}(1+\omega_{\nu})=\beta\rho_{\nu}(1-3\omega_{\nu})(\dot{\sigma}+\dot{\phi})
	\end{split}
\end{equation}

and by the Klein-Gordon equations we have
\begin{equation}\label{fried}
	\begin{split}
		\ddot{\phi}=\lambda_{\phi}V-\frac{3}{2}H\dot{\phi}(1+\omega_{\phi})-\frac{3HV}{\dot{\phi}}(1+\omega_{\phi})
		-\\ \beta\rho_{\nu}(1-3\omega_{\nu})
	\end{split}
\end{equation}

\begin{equation}\label{fried}
	\begin{split}
		\ddot{\sigma}=-\lambda_{\sigma}V-\frac{3}{2}H\dot{\sigma}(1+\omega_{\sigma})+\frac{3HV}{\dot{\sigma}}(1+\omega_{\sigma})
		\\+\beta\rho_{\nu}(1-3\omega_{\nu})
	\end{split}
\end{equation}

To recreate them in a dynamic system, we define the EN variables
\begin{equation}\label{fried}
	\begin{split}
		\eta_{1}=\frac{\kappa^{2}\rho_{\rm b}}{3H^{2}} \ \ \ \ \eta_{2}=\frac{\kappa^{2}\rho_{\rm c}}{3H^{2}} \ \ \ \ \eta_{3}=\frac{\kappa^{2}\rho_{\rm r}}{3H^{2}} \ \ \ \ \eta_{4}=\frac{\kappa^{2}\rho_{\rm \nu}}{3H^{2}} \ \ \ \\ \                                            \eta_{5}=\frac{\kappa\dot{\phi}}{\sqrt{6}H} \ \ \ \eta_{6}=-\frac{\kappa\dot{\sigma}}{\sqrt{6}H} \ \ \
		\eta_{7}=\frac{\kappa^{2}V(\sigma,\phi)}{3H^{2}}
	\end{split}
\end{equation}
To find the dynamics of the system governing cosmic evolution, we follow the same method we used for the previous two models.
\begin{equation}\label{fried}
	\frac{d\eta_{1}}{dN}=-3\eta_{1}-2\eta_{1}\frac{\dot{H}}{H^{2}}
\end{equation}
\begin{equation}\label{fried}
	\frac{d\eta_{3}}{dN}=-4\eta_{3}-2\eta_{3}\frac{\dot{H}}{H^{2}}
\end{equation}
\begin{equation}\label{fried}
	\begin{split}
	\frac{d\eta_{4}}{dN}=-3\eta_{4}(1+\omega_{\nu})- \beta\sqrt{6}(1-3\omega_{\nu})\eta_{4}\eta_{6}
\\	+\beta\sqrt{6}(1-3\omega_{\nu})\eta_{4}\eta_{5}
	-2\eta_{4}\frac{\dot{H}}{H^{2}}
	\end{split}
\end{equation}

\begin{equation}\label{fried}
	\begin{split}
	\frac{d\eta_{5}}{dN}=\frac{3\lambda_{\phi}}{\sqrt{6}}\eta_{7}-\frac{3}{2}(1+\omega_{\phi})\eta_{5}-
	\frac{3}{2}(1+\omega_{\phi})\frac{\eta_{7}}{\eta_{5}}
\\	-\frac{3\beta}{\sqrt{6}}(1-3\omega_{\nu})\eta_{4}-\eta_{5}\frac{\dot{H}}{H^{2}}
	\end{split}
\end{equation}

\begin{equation}\label{fried}
	\begin{split}
	\frac{d\eta_{6}}{dN}=\frac{3\lambda_{\sigma}}{\sqrt{6}}\eta_{7}-\frac{3}{2}(1+\omega_{\sigma})\eta_{6}-
	\frac{3}{2}(1+\omega_{\phi})\frac{\eta_{7}}{\eta_{6}}
	\\-\frac{3\beta}{\sqrt{6}}(1-3\omega_{\nu})\eta_{4}+\eta_{6}\frac{\dot{H}}{H^{2}}
		\end{split}
\end{equation}
\begin{equation}\label{fried}
	\frac{d\eta_{7}}{dN}=\eta_{7}(\sqrt{6}\lambda_{\sigma}\eta_{6}-\sqrt{6}\lambda_{\phi}\eta_{5}-2\frac{\dot{H}}{H^{2}})
\end{equation}

and the Friedmann constraint is

\begin{equation}\label{fried}
	\begin{split}
		\eta_{2}=1-\eta_{1}-\eta_{3}-\eta_{4}-\eta_{5}^{2}+\eta_{6}^{2}-\eta_{7}
	\end{split}
\end{equation}
holds.

\color{black}
\section{Numerical Analysis}
Connecting theory to data is an integral part of the scientific process. Generally a cosmological observable refers to a specific phenomenon or class of objects which can be used
to measure some key properties of cosmology, most commonly the clustering of matter
at some particular epoch and within some set of length scales. These observations allow
us to compare theoretical predictions to data, and thereby  constraint properties of the
cosmological model. All observational data  used in this paper are:\\
$\bullet$ Pantheon catalog:
\cite{Scolnic} compiled the Pantheon sample consisting 1701 SNe Ia covering the redshift range $0.001 < z < 2.3$.\\
$\bullet$ {CMB data}:
We used the latest large-scale cosmic microwave background (CMB) temperature and
polarization angular power spectra from the final release of Planck 2018 plikTTTEEE+lowl+lowE
\cite{Aghanim}. \\
$\bullet$ {BAO data}:
We also used the various measurements of the Baryon Acoustic Oscillations (BAO) from
different galaxy surveys \cite{Aghanim}, i.e.
6dFGS.(2011)\cite{Beutler}, SDSS-MGS
\cite{Ross}, and BOSS DR12 (2017)\cite{Alam}.

To analyze the data and extract the constraints on these cosmological parameters, we
used our modiﬁed version of the publicly available Monte Carlo Markov Chain package
CosmoMC \cite{Lewis}. This is equipped with a convergence diagnostic based on the Gelman and
Rubin statistic \cite{Gelman}, assuming $R - 1 < 0.02$, and implements an efﬁcient sampling of the pos-
terior distribution using the fast/slow parameter decorrelations \cite{Lewis1}. CosmoMC includes
support for the 2018 Planck data release \cite{Aghanim}. We also used the Akaike Information Criteria (AIC)
\begin{equation}\label{key}
	AIC = \chi _{\min }^2 + 2k	
\end{equation}
In these equations$\chi _{\min }^2  $ is the minimum value of $ {\chi ^2} $, k is the number of parameters of the given model. AIC provides means to
compare models with different numbers of parameters; they
penalize models with a higher k in favor of those with a lower
k , in effect enforcing Occam’s Razor in the model selection
process.

We put constraints on the following cosmological parameters: the baryon energy density ${\Omega _b}{h^2}$, the cold dark matter energy density $\Omega_{c}h^{2}$, ($\sum m_{\nu}$ total mass of neutrino, $N_{\rm eff}$ effective number of relativistic species, $H_{0}$ Hubble constant, $\lambda$ a dimensionless parameter that determines the slope of the potential.\\
$\bullet$$\bullet$ Phantom Model:\\
In what follows, we put constraint on the total mass of neutrino by analyzing of Pantheon and CMB and BAO data. We first survey the results of the CMB + Pantheon data and then investigate  the results of CMB + BAO, and finally survey the total results of these two parts. The results for the cosmological parameters  are shown
in Table I. Fig. 1 also shows the parametric
space at 68 $\%$CL and 95$\%$CL for some selected parameters for the different observational data
sets.
\\

From the analysis of the CMB + BAO data and, for phantom we find that
$\sum m_{\nu}<0.168$eV \ \ (95$\% $CL.)  and using CMB+Pantheon+ we find $ \sum m_{\nu}<0.22$eV \ \ (95$\% $CL.)  and for combination of full data(CMB+BAO+Pantheon+) we find $ \sum m_{\nu}<0.121$eV \ \ (95$\% $CL.).\\
For results from  combined data(Pantheon+CMB+BAO), we consider multivariate joint Gaussian likelihood given by
\begin{equation}\label{27}
	\mathcal{L}_{\rm Joint} \propto \text{exp}\left(\frac{-\chi^2_{\rm Joint}}{2}\right),
\end{equation}
where the joint chi-squared function of all the datasets reads
\begin{equation}
	\chi^2_{\rm Joint} = \chi^2_{\rm BAO} +\chi^2_{\rm CMB} + \chi^2_{\rm Pantheon}.
\end{equation}
\begin{figure*}
	\centering
	\includegraphics[scale=.4]{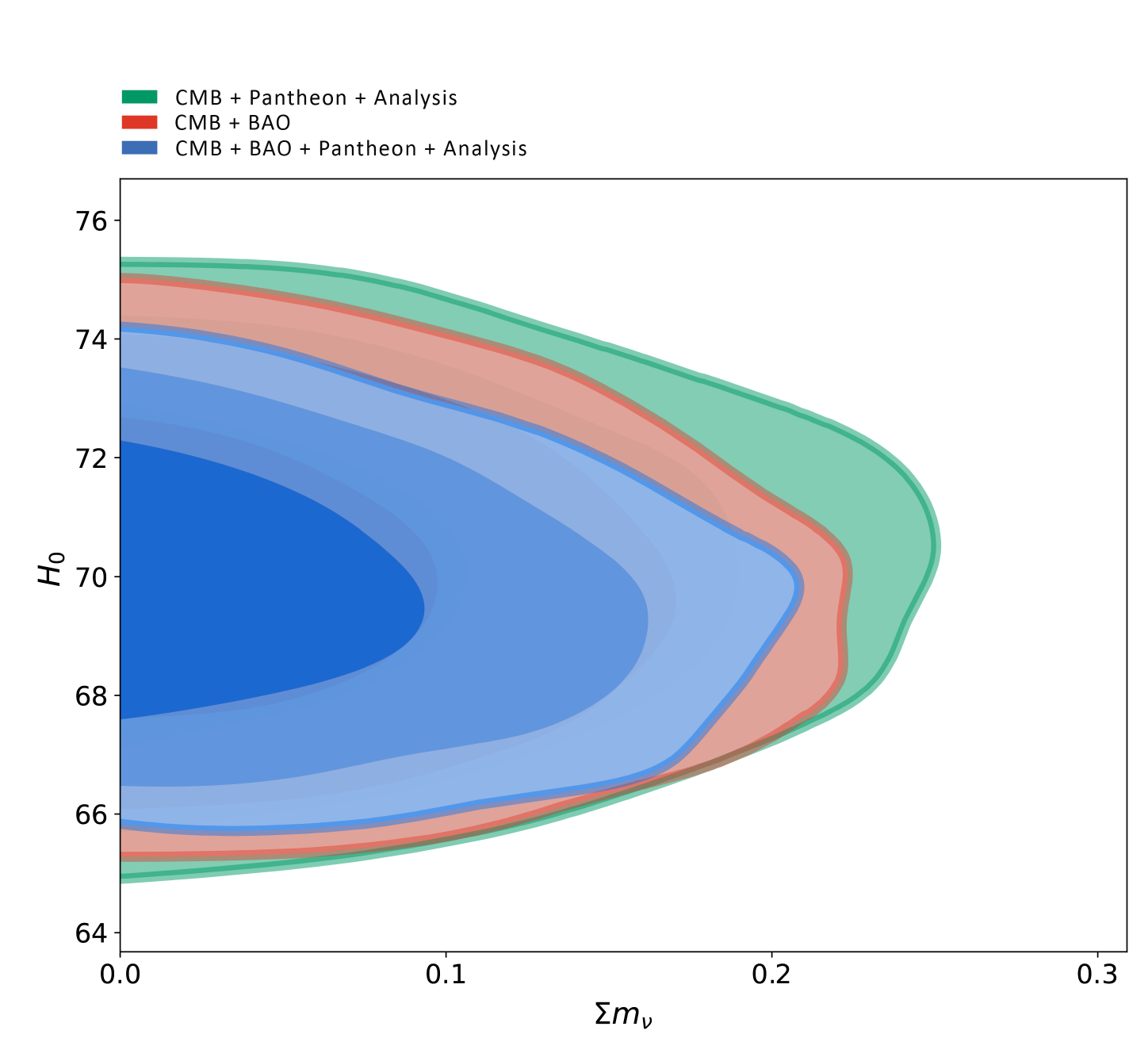}\hspace{0.05 cm}\\
	Figure 1: The constraints at the (95$\% $CL.) two-dimensional contours for $\sum m_{\nu}$ for phantom model
\end{figure*}
\begin{figure*}
	\centering
	\includegraphics[scale=.4]{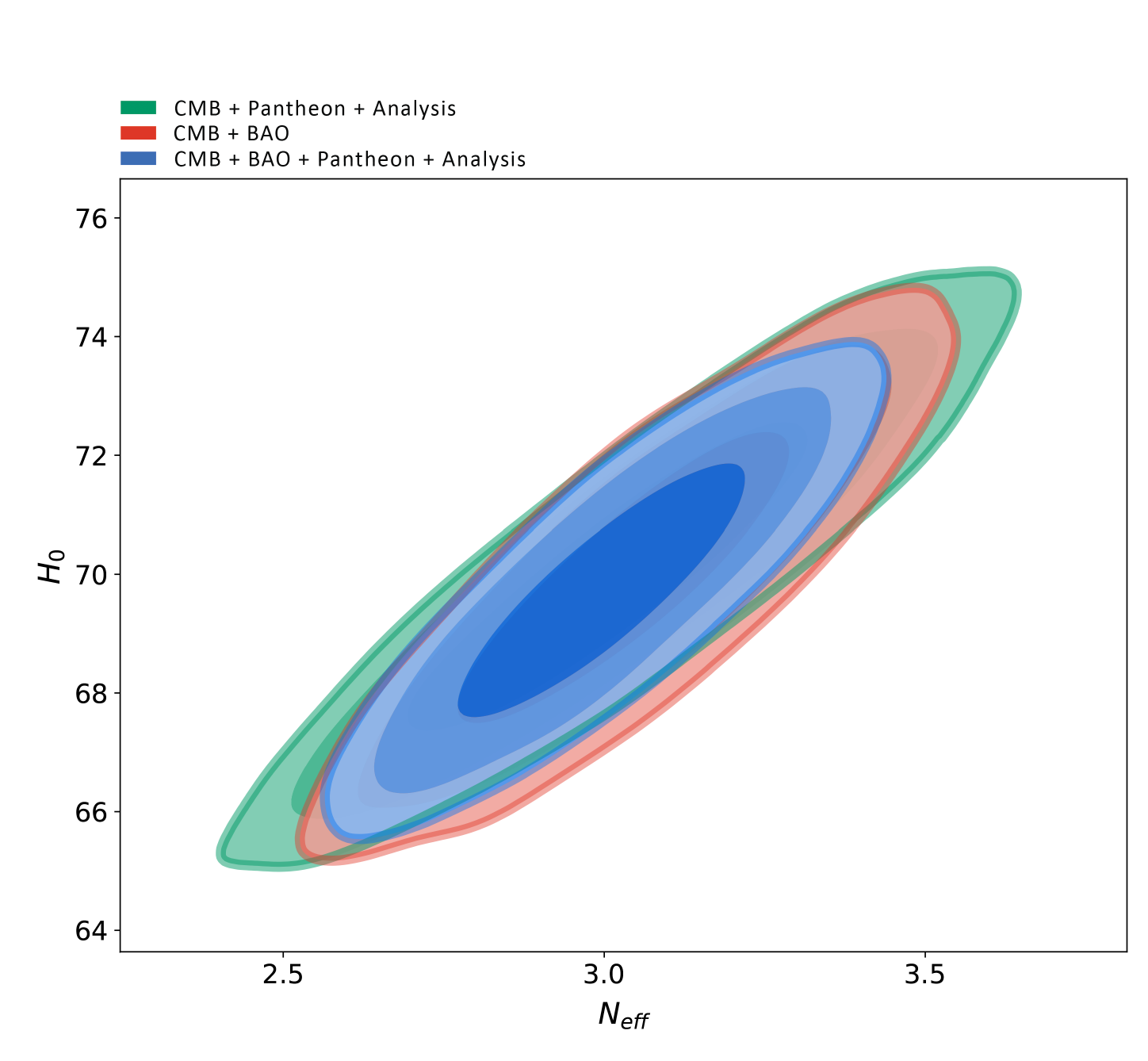}\hspace{0.05 cm}\\
	Figure 2: The constraints at the (68$\% $CL.) two-dimensional contours for $N_{\rm eff}$ in  phantom model.
\end{figure*}

\begin{table*}
	\scriptsize
	\caption{ Observational constraints at (95$\% $CL.) on main and derived parameters of the  $\sum m_{\nu}$ scenario. The parameter
		$H_{0}$ is in the units of $km/sec/Mpc$, whereas $\sum m_{\nu}$ reported in the (95$\% $CL.), is in the units of eV (Phantom model).
	} % title of Table
	\centering % used for centering table
	\begin{tabular}{c@{\hspace{3mm}} c@{\hspace{1mm}} c@{\hspace{1mm}} c@{\hspace{1mm}} c@{\hspace{1mm}}
			c@{\hspace{1mm}} c@{\hspace{1mm}} c@{\hspace{3mm}} c@{\hspace{1mm}}  c@{\hspace{1mm}}
			c@{\hspace{1mm}} c@{\hspace{1mm}}c@{\hspace{1mm}}c@{\hspace{1mm}}c@{\hspace{1mm}}} % centered columns (5 columns)
		\hline\hline %inserts double horizontal lines
		Dataset  &  $\Omega_{\rm b}h^{2}$  & $\Omega_{\rm c}h^{2}$  & $H_{0}$ & $\Omega_{\rm m}$ & $\Omega_{\rm m}h^{2}$ & $\sum m_{\nu}$ & $\beta$ & $\lambda$& $N_{\rm eff}$ &\\% inserts table
		\hline % inserts single horizontal line
		CMB + BAO & $0.02233\pm 0.00019$ & $0.1184\pm 0.0034 $ &$69.91\pm 1.4 $ & $0.3087\pm 0.0071$ & $0.1414^{+0.0024}_{-0.0021}$ & $<0.168$ & $0.173^{+0.37}_{-0.25}$ & $1.01\pm 0.4$ & $2.98\pm 0.17$ & \\% inserts table
		%heading
		\hline % inserts single horizontal line
		CMB+Pantheon& $0.02238\pm 0.00018\pm 0.00019$ & $0.1181\pm 0.0029$ & $71.05\pm1$ & $0.3017\pm 0.0071 $ & $0.1272\pm0.004$ & $<0.22$ & $0.171\pm0.29$ & $0.72^{+0.2}_{-0.3}$ & $3.02\pm 0.20$ & \\% inserts table
		%heading
		\hline % inserts single horizontal line
		CMB+BAO+Pantheon& $0.02237\pm 0.00018$ & $0.1187\pm 0.0025$ & $70.4\pm2$ & $0.3111\pm 0.0071 $ & $0.1263\pm0.002$ & $<0.121$ & $0.229\pm0.42$ & $0.8^{+0.5}_{-0.5}$ & $3.00\pm 0.15 $ & \\% inserts table
		%heading
		\hline % inserts single horizontal line
	\end{tabular}
\end{table*}

$\bullet$$\bullet$ Quintessence Model:\\
From the analysis of the CMB + BAO data , for quintessence we find that
$\sum m_{\nu}<0.162$eV \ \ (95$\%$ CL.) and using CMB+Pantheon+ we find $ \sum m_{\nu}<0.224$eV \ \ (95$\% $CL.)  and for combination of full data(CMB+BAO+Pantheon+) we find $ \sum m_{\nu}<0.122$eV \ \ (95$\%$ CL.).

\begin{figure*}
	\centering
	\includegraphics[scale=.4]{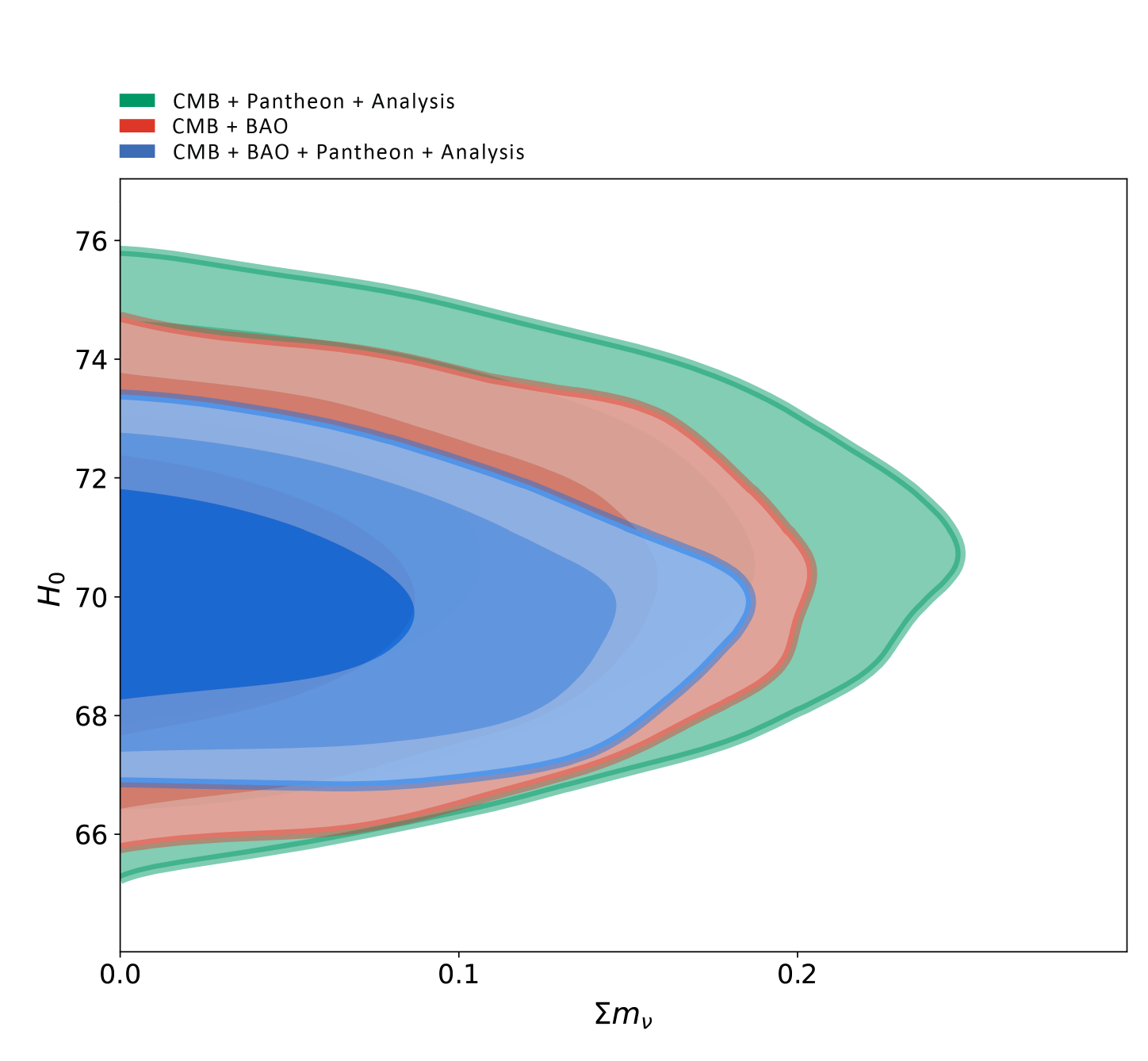}\hspace{0.1 cm}\\
	Figure 3: The constraints at the (95$\% $CL.) two-dimensional contours for $\sum m_{\nu}$ in Quintessence model.
\end{figure*}
\begin{figure*}
	\centering
	\includegraphics[scale=0.4]{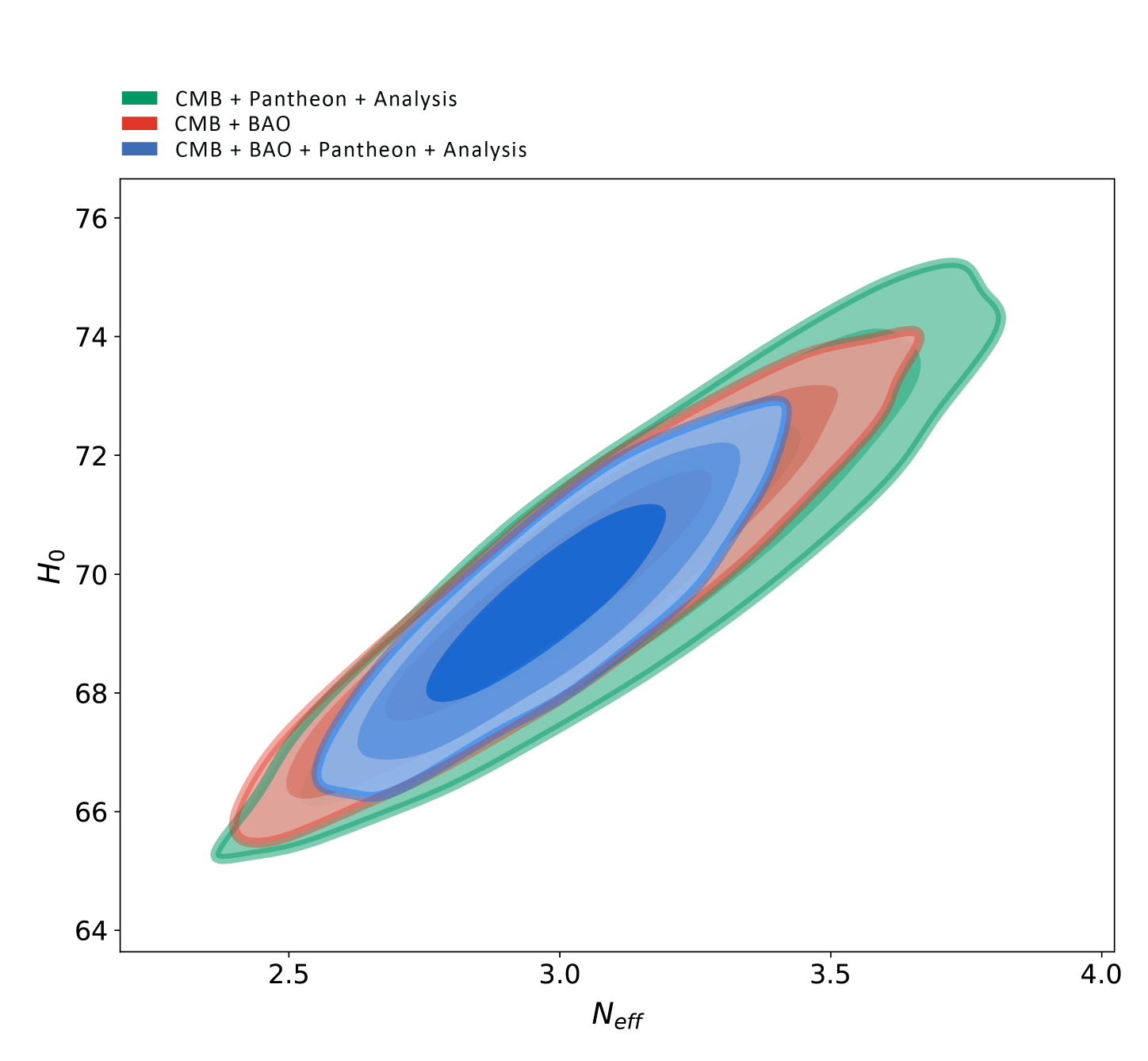}\hspace{0.1 cm}\\
	Figure 4: The constraints at the (68$\% $CL.) two-dimensional contours for $N_{\rm eff}$ in  Quintessence model.
\end{figure*}

\begin{table*}
	\scriptsize	
	\caption{: Observational constraints at (95$\% $CL.) on main and derived parameters of the  $\sum m_{\nu}$ scenario. The parameter
		$H_{0}$ is in the units of $km/sec/Mpc$, whereas $\sum m_{\nu}$ reported in the $95\%$ CL, is in the units of eV (Quintessence Model).
	} % title of Table
	\centering % used for centering table
	\begin{tabular}{c@{\hspace{2mm}} c@{\hspace{2mm}} c@{\hspace{2mm}} c@{\hspace{2mm}} c@{\hspace{2mm}}
			c@{\hspace{2mm}} c@{\hspace{2mm}} c@{\hspace{2mm}} c@{\hspace{2mm}}  c@{\hspace{2mm}}
			c@{\hspace{2mm}} c@{\hspace{2mm}}c@{\hspace{2mm}}c@{\hspace{2mm}}c@{\hspace{2mm}}} % centered columns (5 columns)
		\hline\hline %inserts double horizontal lines
		Dataset  &  $\Omega_{\rm b}h^{2}$  & $\Omega_{\rm c}h^{2}$  & $H_{0}$ & $\Omega_{\rm m}$ & $\Omega_{\rm m}h^{2}$ & $\sum m_{\nu}$ & $\beta$ & $\lambda$& $N_{\rm eff}$ &\\% inserts table
		%heading
		\hline % inserts single horizontal line
		CMB+BAO & $0.02231\pm 0.00019$ & $0.1181^{+0.0030}_{-0.0034}$ & $69.88\pm 1.8 $ & $0.3092\pm 0.0084$ & $0.127^{+0.0028}_{-0.0027}$ & $<0.162$ & $0.186^{+0.31}_{-0.36}$ & $0.22\pm0.3$ & $2.99\pm 0.20 $ & \\% inserts table
		%heading
		\hline % inserts single horizontal line
		CMB+Pantheon& $0.02225\pm 0.00024$ & $0.1200\pm 0.0037 $ & $70.54\pm1.4$ & $0.325\pm0.01$ & $0.1263\pm0.004$ & $<0.324$ & $0.141\pm0.29$ & $0.714^{+0.4}_{-0.35}$ & $3.11\pm 0.23  $ & \\% inserts table
		%heading
		\hline % inserts single horizontal line
		CMB+BAO+Pantheon& $0.02236\pm 0.00017 $ & $0.1183\pm 0.0024$ & $70.75\pm1.23$ & $0.3108\pm 0.0069  $ & $0.124^{+0.006}_{-0.005}$ & $<0.122$ & $0.632^{+0.7}_{-0.7}$ & $8.1^{+0.74}_{-0.63}$ & $2.98\pm 0.14 $ & \\% inserts table
		%heading
		\hline % inserts single horizontal line
	\end{tabular}
\end{table*}

$\bullet$$\bullet$ Quintom Model:\\

For CMB + BAO data, we find that
$\sum m_{\nu}<0.128$eV \ \ (95$\% $CL.)  and using CMB+Pantheon+ we find $ \sum m_{\nu}<0.2$eV \ \ (95$\% $CL.)  and for combination of full data(CMB+BAO+Pantheon+) we find $ \sum m_{\nu}<0.1197 $eV \ \ (95$\% $CL.)

\begin{figure*}
	\centering
	\includegraphics[scale=.4]{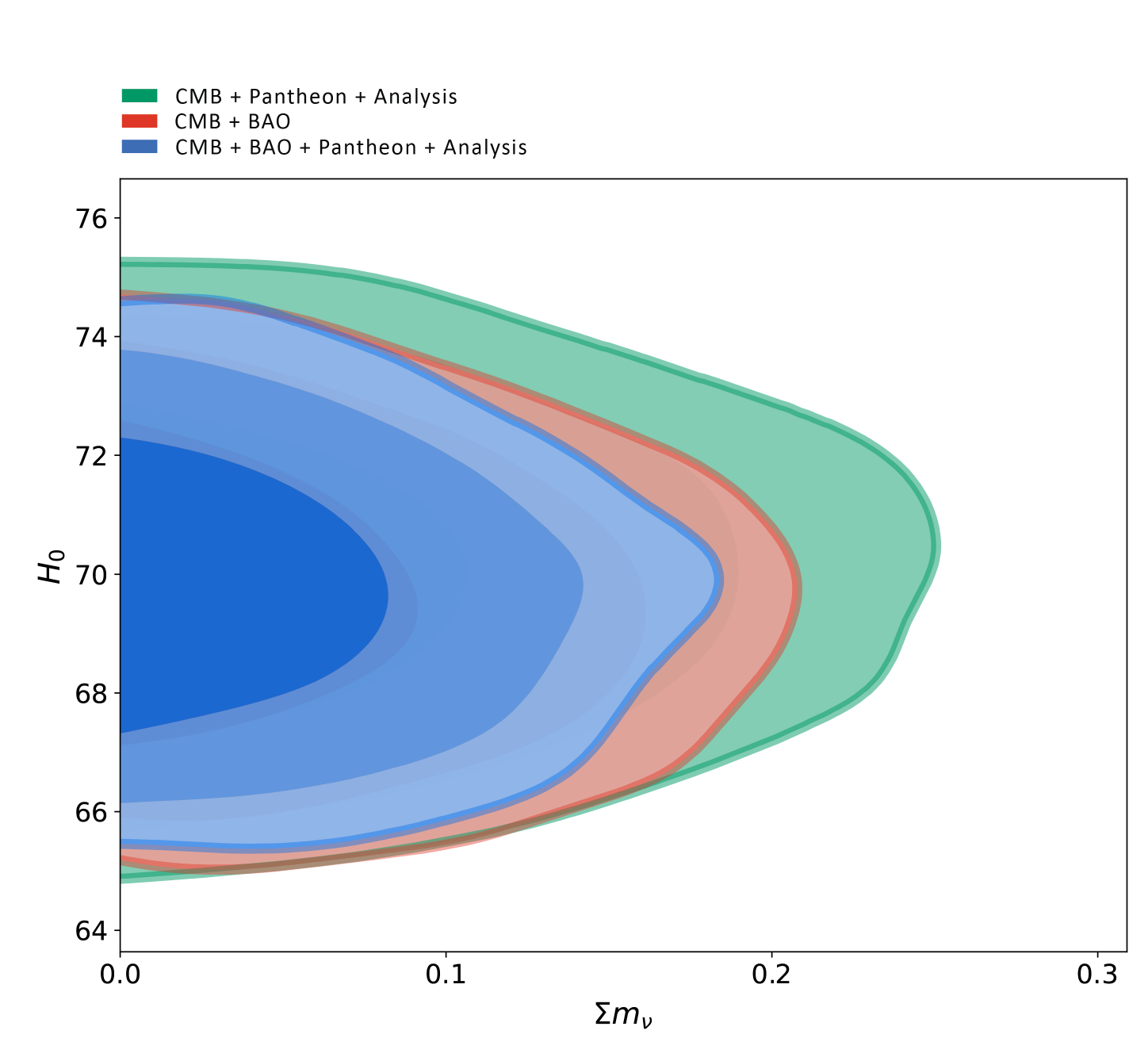}\hspace{0.1 cm}\\
	Figure 5: The constraints at the (95$\% $CL.) two-dimensional contours for $\sum m_{\nu}$ in Quintom model.
\end{figure*}

\begin{figure*}
	\centering
	\includegraphics[scale=.4]{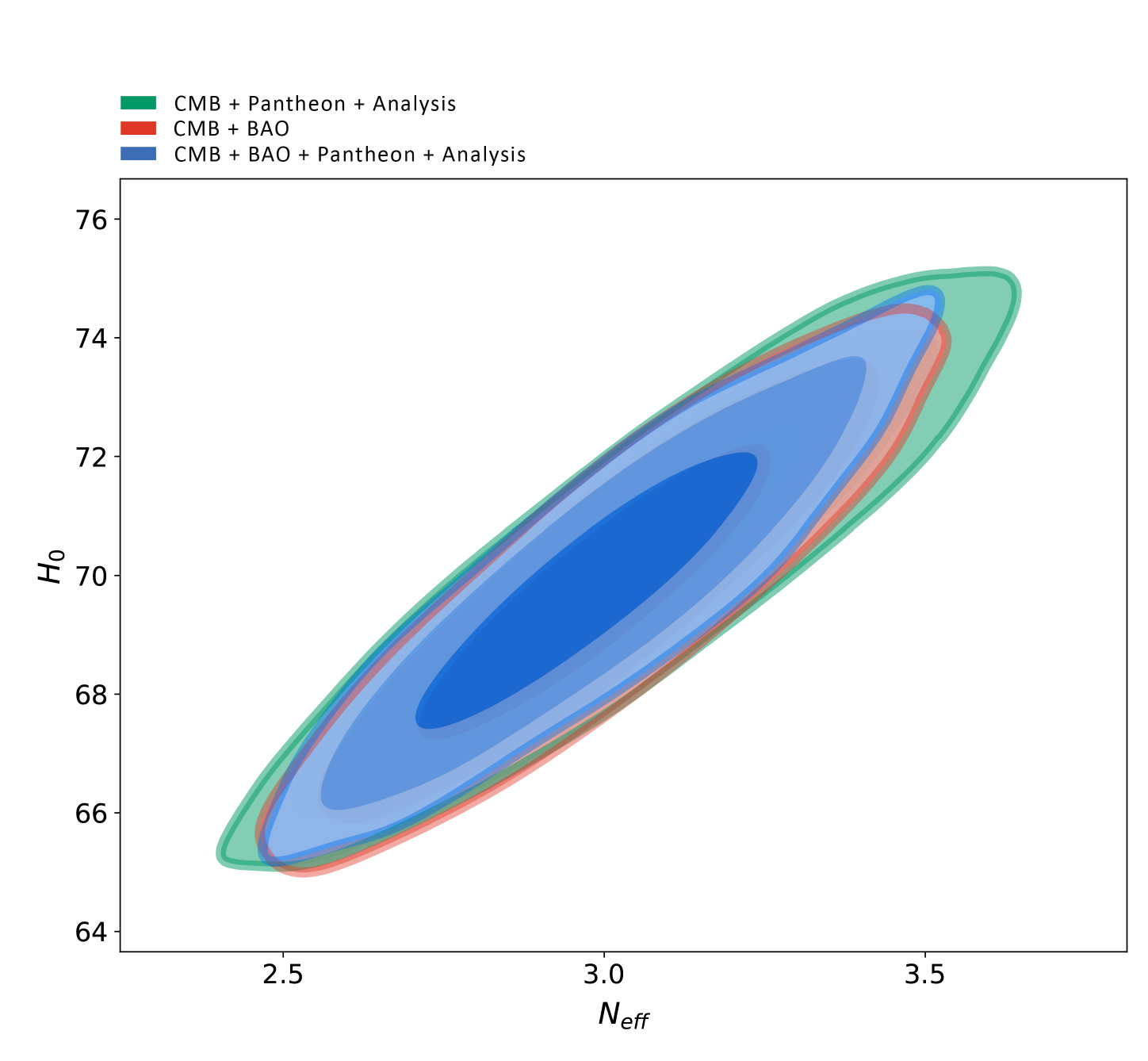}\hspace{0.1 cm}\\
	Figure 6: The constraints at the (68$\% $CL.) two-dimensional contours for $N_{\rm eff}$ in quintom model.
\end{figure*}

This result is close to results of  \cite{Aghanim} TT,TE,EE +lowE+lensing+BAO with $\sum m_{\nu}<0.12$eV at 95$\% $ CL and case TT,TE,EE+lowE+BAO with $\sum m_{\nu}<0.13$ eV at 95$\%$ CL
in three models.
Other parameters are shown in table III.\\

\begin{table*}
	\scriptsize	
	\caption{: Observational constraints at (95$\% $CL.) on main and derived parameters of the  $\sum m_{\nu}$ scenario. The parameter
		$H_{0}$ is in the units of $km/sec/Mpc$, whereas $\sum m_{\nu}$ reported in the (95$\% $CL.), is in the units of (eV) (Quintom).
	} % title of Table
	\centering % used for centering table
	\begin{tabular}{c@{\hspace{3mm}} c@{\hspace{3mm}} c@{\hspace{3mm}} c@{\hspace{3mm}} c@{\hspace{3mm}}
			c@{\hspace{1mm}} c@{\hspace{1mm}} c@{\hspace{1mm}} c@{\hspace{1mm}}  c@{\hspace{1mm}}
			c@{\hspace{1mm}} c@{\hspace{1mm}}c@{\hspace{1mm}}c@{\hspace{1mm}}c@{\hspace{1mm}}} % centered columns (5 columns)
		\hline\hline %inserts double horizontal lines
		Dataset  &  $\Omega_{\rm b}h^{2}$  & $\Omega_{\rm c}h^{2}$  & $H_{0}$ & $\Omega_{\rm m}$  & $\sum m_{\nu}$ & $\beta$ & $\lambda_{\sigma}$&$\lambda_{\phi}$& $N_{\rm eff}$ &\\% inserts table
		%heading
		\hline % inserts single horizontal line
	CMB+BAO	 & $0.02238\pm 0.00019$ & $0.1185\pm 0.0030 $ & $69.95\pm 1.73 $ & $0.3109\pm 0.0073$ &  $<0.128$ & $0.151\pm0.25$ & $-2.75\pm0.5$ & $2.83\pm0.07$ & $2.99\pm 0.18$ & \\% inserts table
		%heading
		\hline % inserts single horizontal line
		CMB+Pantheon & $0.02233\pm 0.00017 $ & $0.1181\pm 0.0034  $ & $70.92\pm1.55$ & $0.3087\pm 0.0071 $ & $<0.2$ & $0.164\pm0.4$ & $-2.9\pm0.05$ &$1.91^{+0.011}_{-0.017}$ & $2.97^{+0.32}_{-0.35}$ & \\% inserts table
		%heading
		\hline % inserts single horizontal line
		CMB+BAO+Pantheon& $0.02238\pm 0.00012   $ & $0.1180\pm 0.0019 $ & $71.03\pm1.1$ & $0.3099\pm 0.0054$ &  $<0.119$ & $0.236\pm0.4$ & $-2.63^{+0.05}_{-0.03}$ &$2.6^{+0.014}_{-0.028}$ & $2.98\pm 0.17 $ & \\% inserts table
		%heading
		\hline % inserts single horizontal line
	\end{tabular}
\end{table*}

\begin{table*}
	\caption{{\small  Mean values of free parameters of various models with 1$ \sigma $ error bar for combinations data}}
	\begin{center}
		\resizebox{1\textwidth}{!}{  
			\begin{tabular}{ c |c c c c c c c c c  } 
				\hline
				\hline
				Models & $ \Omega_{\phi}$ & $ \Omega_{\rm m}$  & $ \Omega_{\Lambda}$&$ \Omega_{\sigma}$&$\beta$&$ \Omega_{\nu}$&$ H_{0} $& AIC \\ 
				\hline
				$\Lambda$CDM  & $-$ & $0.311\pm0.15$ & $0.678\pm0.17$ & $-$ & $-$& $-$ & $68.9\pm1.3$ & $3599.172$   \\
				Phantom& $-$ & $0.309\pm 0.0055$ & $-$ & $0.61\pm0.075$ & $0.229\pm0.43$& $0.0012\pm 0.0009$ & $70.04\pm 1.42$ & $ 3598.873 $  \\
				Quintessence & $0.64\pm0.04$ & $0.3108\pm0.0069$ & $-$ & $-$ & $0.63\pm0.7$& $0.003\pm0.0014$ & $70.25\pm1.23$ & $3597.766$   \\
				Quintom & $0.08\pm0.022$ & $0.3111\pm 0.0071$ & $-$ & $0.62\pm0.064$ & $0.236\pm0.4$ & $0.0026\pm0.0011$ & $71.03\pm1.1$ & $3593.151$  \\
				\hline
				\hline
			\end{tabular}
		}
	\end{center}
	\label{table_chi}
\end{table*}

\begin{table*}
	\caption{{\small $\chi^2$s comparison between $\Lambda$CDM and Phantom ,Quintessence, and Quintom for the different dataset combinations explored in this work.}}
	\begin{center}
		\resizebox{0.6\textwidth}{!}{  
			\begin{tabular}{ c |c c c  } 
				\hline
				\hline
				$\Lambda$CDM  & CMB+Pantheon+ & CMB+BAO & CMB+BAO+Pantheon+  \\ 
				\hline
				$\chi^2_{\rm tot}$ & $3584.778$ & $2776.072$ & $3593.172$  \\
				$\chi^2_{\rm  CMB}$ & $2770.254$ & $2769.860 $ & $2771.442$  \\
				$\chi^2_{\rm  BAO}$ & $-$ & $ 6.212 $ & $6.754$ \\
				$\chi^2_{\rm  Pantheon}$ & $ 814.524 $ & $-$ & $ 814.976 $  \\
				\hline
				\hline
				Phantom   & CMB+Pantheon+ & CMB+BAO & CMB+BAO+Pantheon+\\ 
				\hline
				$\chi^2_{\rm tot}$ & $3582.5$ & $2768.279$ & $3588.873$  \\
			$\chi^2_{\rm  CMB}$ & $2769.237$ & $2762.864 $ & $2770.013$  \\
			$\chi^2_{\rm  BAO}$ & $-$ & $ 5.415 $ & $5.752$ \\
			$\chi^2_{\rm  Pantheon}$ & $ 813.263 $ & $-$ & $ 813.108 $ \\

				\hline
				\hline
				Quintessence  & CMB+Pantheon+ & CMB+BAO & CMB+BAO+Pantheon+ \\ 
				\hline
				$\chi^2_{\rm tot}$ & $3580.152$ & $2766.577$ & $3587.766$  \\
			$\chi^2_{\rm  CMB}$ & $2766.146$ & $2761.479 $ & $2768.365$  \\
			$\chi^2_{\rm  BAO}$ & $-$ & $ 5.098 $ & $5.176$ \\
			$\chi^2_{\rm  Pantheon}$ & $ 814.006 $ & $-$ & $ 814.225 $ \\
				
				\hline
				\hline
				Quintom  & CMB+Pantheon+ & CMB+BAO & CMB+BAO+Pantheon+ \\ 
				\hline
			$\chi^2_{\rm tot}$ & $3580.991$ & $2767.467$ & $3583.151$  \\
			$\chi^2_{\rm  CMB}$ & $2767.821$ & $2762.113 $ & $2769.754$  \\
			$\chi^2_{\rm  BAO}$ & $-$ & $ 5.354 $ & $5.288$ \\
			$\chi^2_{\rm  Pantheon}$ & $ 813.991 $ & $-$ & $ 813.406 $ \\
				\hline
				\hline
			\end{tabular}
		}
	\end{center}
	\label{table_chi}
\end{table*}

Moreover, we obtain the  $ \Delta $(AIC) Between 	$\Lambda$CDM and each model.\\
$\bullet$ The value of $ \Delta $(AIC) between the phantom model and $\Lambda$CDM  is 0.299\\

$\bullet$  The value of $ \Delta $(AIC) between the Quintessence model and $\Lambda$CDM  is 1.406\\

$\bullet$  The value of $ \Delta $(AIC) between the Quintom model and $\Lambda$CDM  is 6.021\\

\begin{figure*}
	\centering
	\includegraphics[scale=.6]{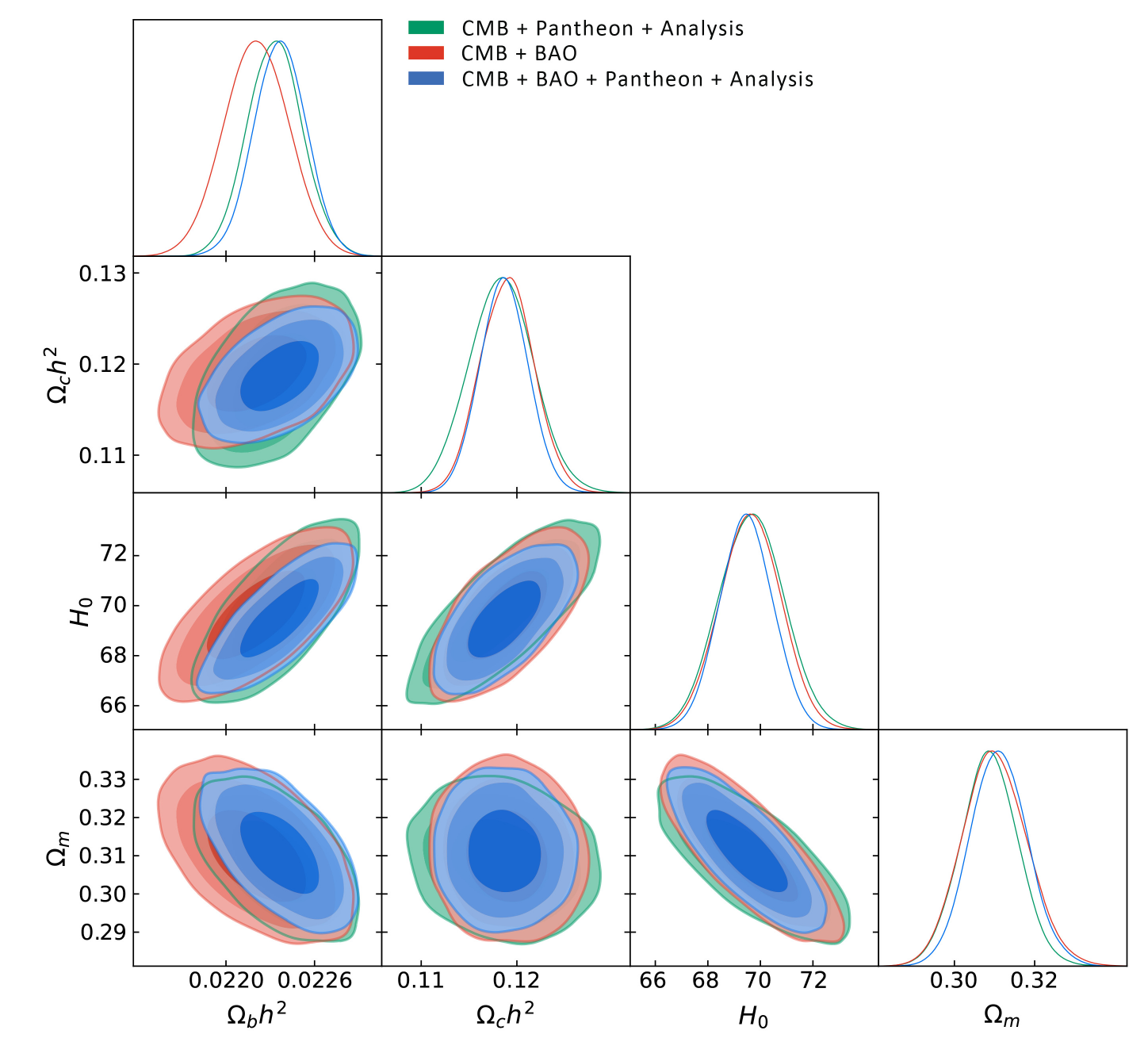}\hspace{0.1 cm}\\
	Figure 7: Comparison of $\Omega_{\rm b}h^{2}$, $\Omega_{\rm c}h^{2}$, $H_{0}$, $\Omega_{\rm m}$ obtained values in Phantom model
\end{figure*}

\begin{figure*}
	\centering
	\includegraphics[scale=.6]{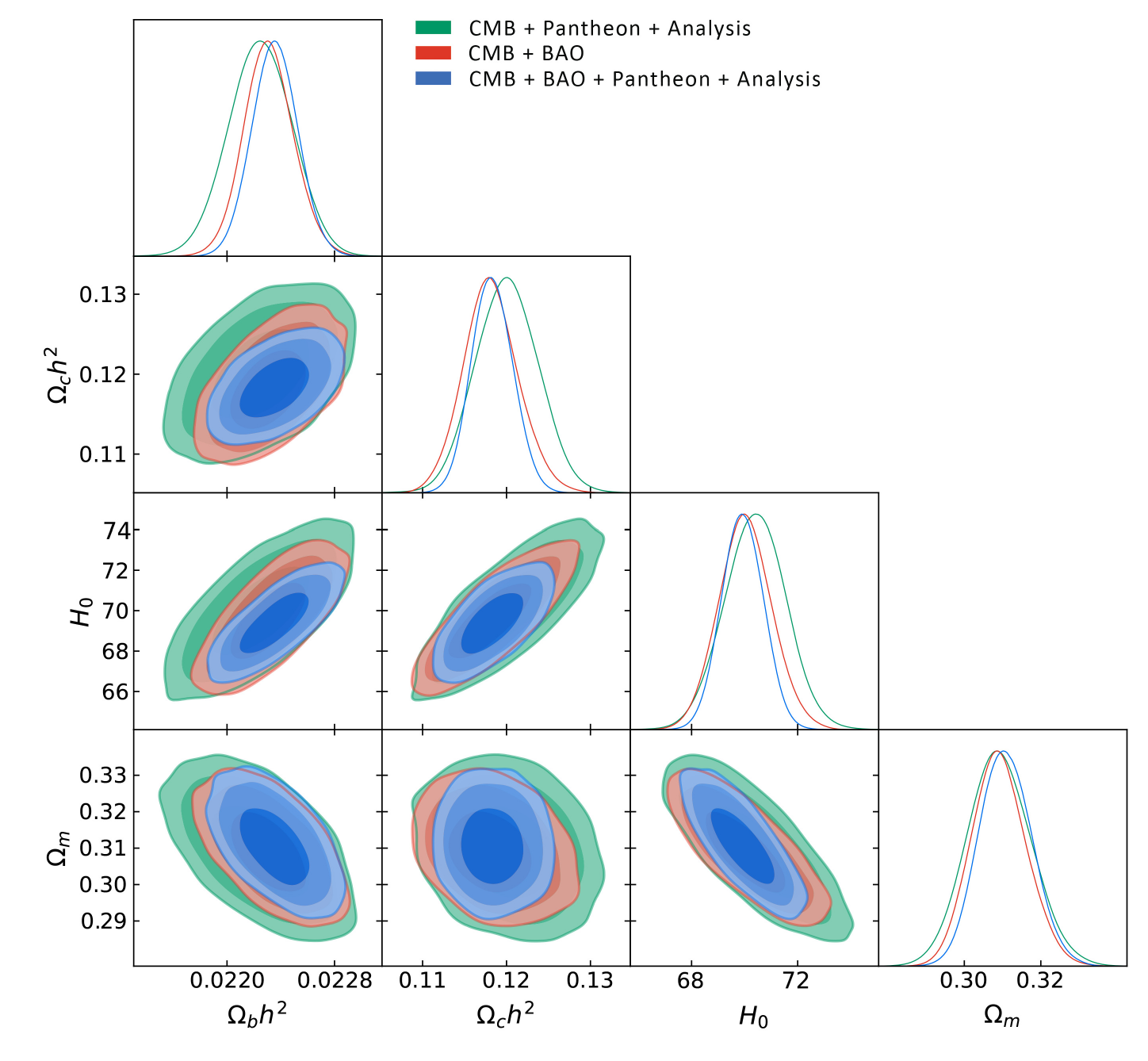}\hspace{0.1 cm}\\
	Figure 8: Comparison of $\Omega_{\rm b}h^{2}$, $\Omega_{\rm c}h^{2}$, $H_{0}$, $\Omega_{\rm m}$ obtained values in Quintessence model
\end{figure*}

\begin{figure*}
	\centering
	\includegraphics[scale=.6]{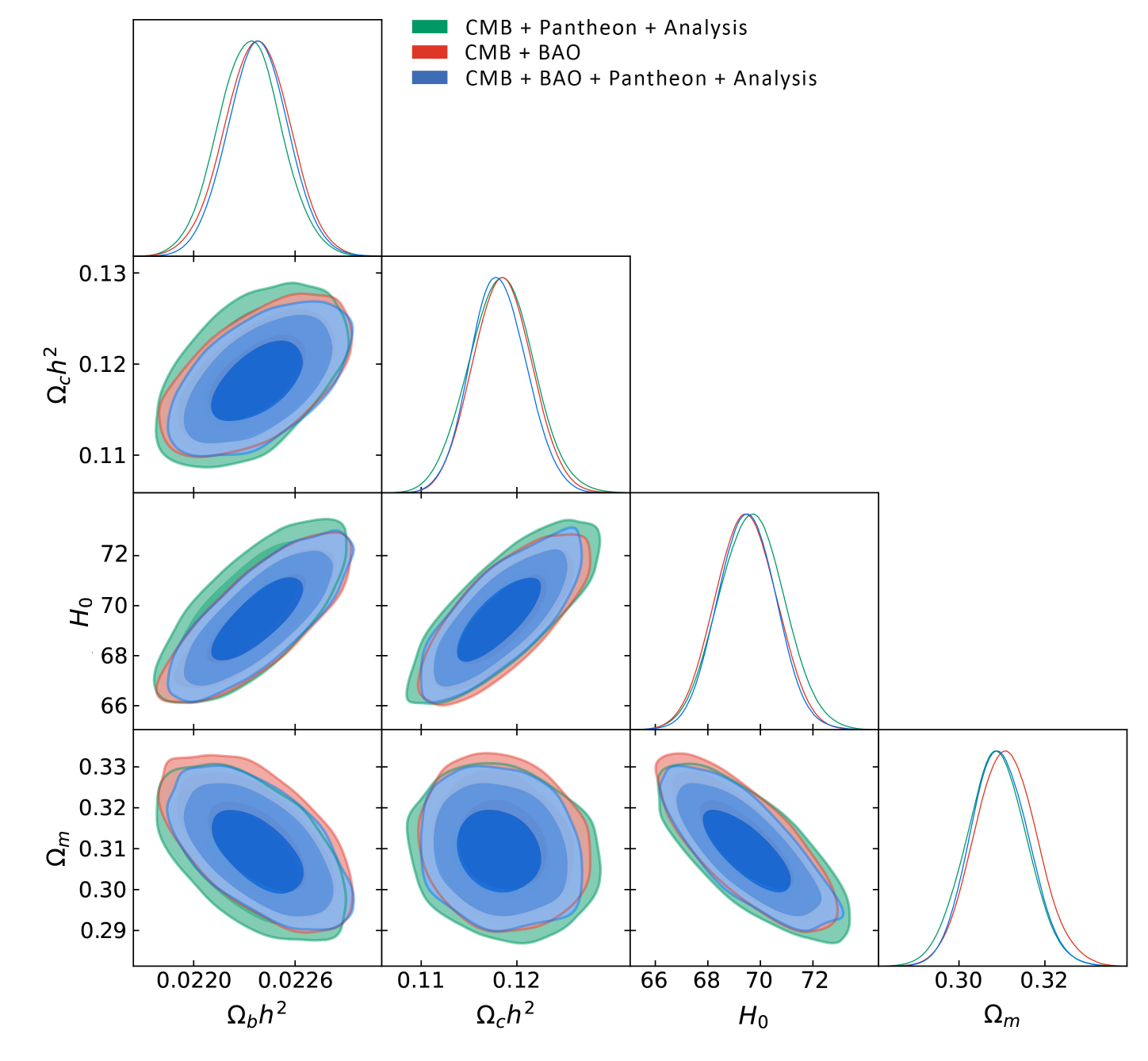}\hspace{0.1 cm}\\
	Figure 9: Comparison of $\Omega_{\rm b}h^{2}$, $\Omega_{\rm c}h^{2}$, $H_{0}$, $\Omega_{\rm m}$ obtained values in Quintom model
\end{figure*}

\begin{figure}
	\centering
	\includegraphics[scale=.5]{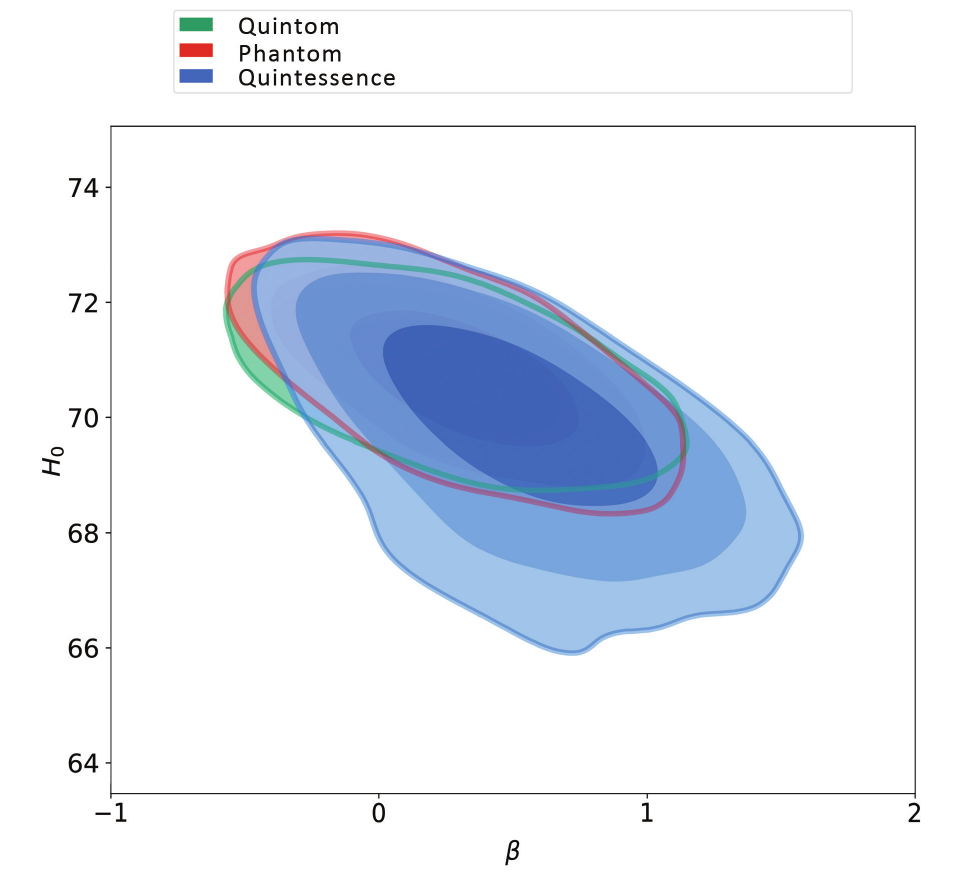}\hspace{0.1 cm}\\
	Figure 10: Comparison of $\beta$ value obtained  for combined data (CMB + BAO + Pantheon) in three  models.
\end{figure}

In the following, we estimate the Neutrinos transition from relativistic to non-relativistic at
redshift $z_{\rm nr}$ for all three models. Neutrinos decouple from the primordial plasma in a Fermi-Dirac distribution:
\begin{equation}
	f({p_\nu },{T_\nu }) = [\exp (\frac{{{p_\nu }}}{{{T_\nu }}}) + 1]^{-1}
\end{equation}
with temperature $T_\nu$. The average momentum is related to the temperature by $\left\langle {{p_\nu }} \right\rangle  = 3.15{T_\nu }$. Massive neutrinos become non-relativistic when $p_\nu $ falls below their rest mass. The temperature of the
CNB is related to the temperature of the CMB by:

\begin{equation}
	{T_\nu }^0 = {(\frac{4}{{11}})^{\frac{1}{3}}}{T^0}_{CMB}
\end{equation}

Using a CMB temperature of $2.725K$ and given that in general $T(z) = {T_0}(1 + z)$, we can then estimate the redshift at which a neutrino of mass $m_{\nu}$ will become non-relativistic as:

\begin{equation}
	{z_{nr}} = (\frac{{{m_\nu }}}{{5.28 \times {{10}^{ - 4}}ev}}) - 1
\end{equation}
According to the above equation and the values obtained for the neutrino mass in all three models, we will calculate the the  redshift at which a neutrino of mass $m_{\nu}$ will become non-relativistic (${z_{nr}}$):\\

$\bullet$  For the phantom model(combination data)  we obtained  ${z_{\rm nr}} = 228.166$

$\bullet$  For the quintessence model(combination data)   we obtained  ${z_{\rm nr}} = 230.060$

$\bullet$  For the quintom model(combination data)  we obtained  ${z_{\rm nr}} = 225.704$\\

Now we can discuss the effect of neutrinos in the structures formation in the early universe.
An important quantity in the context of structure formation is the Jeans length, which traditionally refers to the length scale below which gravitational collapse is counteracted by pressure forces:
\begin{equation}
	{k_J}(t) = {(\frac{{4\pi G\bar \rho (t){a^2}(t)}}{{c_s^2(t)}})^{\frac{1}{2}}}
\end{equation}

Here G is Newton's gravitational constant, $ {\bar \rho (t)}$ is the mean density of the fluid, and ${c_s^2(t)}$
is the squared speed of sound in the fluid. Although massive neutrinos are effectively collisionless particles, replacing ${c_s^2(t)}$ with the thermal velocity,  ${v_{\rm th}}^2(t)$ defines a free-streaming
scale, below which massive neutrinos do not cluster. Using the Friedman equation, we can
then write this as

\begin{equation}
	{k_{FS}}(t) = \sqrt {\frac{3}{2}} \frac{{{v_{\rm th}}(t)}}{{H(t)}}
\end{equation}
For relativistic neutrinos, where $v = c$, this scale is simply the Hubble radius. For non-relativistic neutrinos, their thermal velocity evolves as
\begin{equation}
	{v_{th}} = \frac{{\left\langle {{p_\nu }} \right\rangle }}{{{m_\nu }}} \approx \frac{{3{T_\nu }}}{{{m_\nu }}} = \frac{{3T_\nu ^0}}{{{m_\nu }}}(1 + z) \approx 150(1 + z)(\frac{{1ev}}{{{m_\nu }}})km{s^{ - 1}}
\end{equation}

We can then write an instantaneous free-streaming scale as a function of
neutrino mass:
\begin{equation}
	{k_{FS}}(z) = 0.82\frac{{\sqrt {{\Omega _\Lambda } + {\Omega _{\rm m}}{{(1 + z)}^3}} }}{{{{(1 + z)}^2}}}(\frac{{{m_\nu }}}{{1ev}})\begin{array}{*{20}{c}}
		h&{Mp{c^{ - 1}}}
	\end{array}
\end{equation}

This scale reaches a minimum wavenumber at $z_{\rm nr}$, so we can define a minimum free-
streaming wavenumber:

\begin{equation}
	{k_{\rm nr}} \approx 0.018{\Omega _{\rm m}}^{\frac{1}{2}}(\frac{{{m_\nu }}}{{1ev}})^{\frac{1}{2}}\begin{array}{*{20}{c}}
		h&{Mp{c^{ - 1}}}
	\end{array}
\end{equation}
On scales larger than this, neutrinos effectively evolve like an additional component of
the dark matter. On smaller scales, neutrino free-streaming means the neutrinos do not
cluster, suppressing the matter power spectrum on these scales. In linear theory, this suppression can be analytically calculated to be $(1 - 8{f_{\nu }})$ with respect to the massless
neutrino case on scales $k \gg {k_{\rm nr}}$, where ${f_{\nu }} = \frac{{{\Omega _{\nu} }}}{{{\Omega _{\rm m}}}}$. In turn, free-streaming (non-clustering) neutrinos slow down the growth of gravitational potential wells on scales $\lambda << \lambda_{\rm FS}$ or wave numbers  $k >> k_{\rm FS}$.
What is more, massive neutrinos make up a fraction of the dark matter, however, due to their large thermal velocities, cluster significantly less than cold dark matter (CDM) on small scales.
According to the values obtained for the mass of neutrinos in these three models(quintessence, phantom, and quintom), we will estimate the co-moving wave number.

$\bullet$  For the phantom model(combination data) we obtained the non-relativistic neutrino wavenumber ${k_{\rm nr}}=0.0.000243$Mp${c^{-1}}$

$\bullet$  For the quintessence model(combination data) we obtained the non-relativistic neutrino wavenumber ${k_{\rm nr}}=0.000248$Mp${c^{-1}}$

$\bullet$  For the quintom model(combination data) we obtained the non-relativistic neutrino wavenumber ${k_{\rm nr}}=0.000245$Mp${c^{-1}}$.\\
These results are in general agreement with (\cite{Masatoshi};\cite{Christos}).\\

In addition to the above scenario, we consider the scenario when $\sum m_{\nu}$ equals to the minimum
value in each hierarchy and the lightest neutrino is massless.
\begin{figure}
	\centering
	\includegraphics[scale=.5]{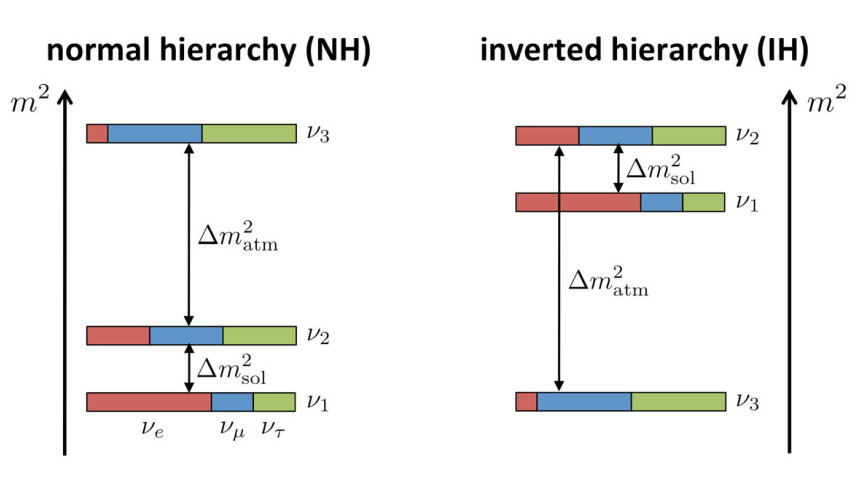}\hspace{0.1 cm}\\
	Figure 11: The results of oscillation experiments lead to two possible orderings of the
	neutrino mass eigenstates.
\end{figure}

The results of oscillation experiments (\cite{Ahmad};\cite{Davis};\cite{Fukuda};\cite{Kajita}) lead to two possible orderings of the
neutrino mass eigenstates. In figure 10: on the left, the normal hierarchy has $ m1 < m2 < m3$ , and
on the right the inverted hierarchy has $m3 > m1 > m2$. The three colors represent the
probability of each mass eigenstate producing a neutrino of each flavor.

The normal
hierarchy has a minimum mass of 0.06 eV, whilst the inverted has a minimum mass of
0.1 eV, meaning a measurement of the neutrino mass scale, $M_\nu = \Sigma_i m_i $ of 0.1 eV to a
high precision would eliminate the possibility of the inverted hierarchy.
We thus have two neutrinos of the same mass
$m_1$ = $ m_2$ and the third one of a different mass $m_3$. We
parameterize the masses in terms of the sum of neutrino
masses  $\sum m_{\nu}$ and the fraction $\theta$ of the total mass in the

third neutrino mass eigenstate, so that
\begin{equation}
	m_{3}=\theta\Sigma m_{i}
\end{equation}
We consider a case where all of the mass is in the third neutrino $(\theta = 1)$ or that
third neutrino is massless $(\theta = 0)$. In the case $(\theta = 1)$, we consider $\sum m_{\nu} = m_{3}$ and another flavors are zero. The result obtained from analysis of this case is the same of results in  above scenario and is close to  inverted hierarchy (minimum mass of
0.1 eV). If we consider the case $(\theta = 0)$, $\sum m_{\nu}$ is split to the two equal masses  and for each model then, we calculate the minimum mass which is very close to the minimum mass of the normal
hierarchy and investigate the effect of non - relativistic  neutrino on the structure formation in early universe.

The results obtained for the normal
hierarchy are:\\
$\bullet$  For the phantom model(combination data)  we obtained  ${z_{\rm nr}} = 113.583$

$\bullet$  For the quintessence model(combination data)   we obtained  ${z_{\rm nr}} = 114.530$

$\bullet$  For the quintom model(combination data)  we obtained  ${z_{\rm nr}} = 112.352$\\
which is in the   matter dominate era. For  non-relativistic neutrino wave number\\
$\bullet$  For the phantom model(combination data) we obtained the non-relativistic neutrino wave number ${k_{\rm nr}}=0.000174$Mp${c^{-1}}$

$\bullet$  For the quintessence model(combination data) we obtained the non-relativistic neutrino wave number ${k_{\rm nr}}=0.000175$Mp${c^{-1}}$

$\bullet$  For the quintom model(combination data) we obtained the non-relativistic neutrino wave number ${k_{\rm nr}}=0.000173$Mp${c^{-1}}$\\

Therefore, before and during the radiation era neutrinos are relativistic and
behave as radiation, while during and after the matter
era neutrinos become non-relativistic and $\omega_{\nu}$ becomes
0. Thus, a complete and detailed investigation of the
thermal history of the universe requires the exact behavior of $\omega_{\nu} (z)$, that is its specific form interpolating between these two regimes. Expressing the universe evolution through the redshift z, for convenience, one can have
several $\omega_{\nu} (z)$ parameterizations with the above required
properties, namely, the interpolation of the equation of
state parameter between $\frac{1}{3}$ to 0.
In this work we consider a semi-relativistic phase (around $z_{nr}$). Following the reference \cite{Wali}, we shall
use the following ansatz for $\omega_{\nu}(z)$
\begin{equation}
	\omega_{\nu}(z)=\frac{p_{\nu}}{\rho_{\nu}}=\left(1+tanh(\frac{ln(1+z)-z_{\rm eq}}{z_{\rm dur}})\right)
\end{equation}
where $z_{\rm eq}$ determines the transition redshift where matter and radiation energy densities become equal and $z_{\rm dur}$ determines how fast this transition is realized. In summary, using the  dimensionless variable $\omega_{\nu}$, we can transform equation (70) into its autonomous form:

\begin{equation}
	\frac{d\omega_{\nu}}{dN}=\frac{2\omega_{\nu}}{z_{\rm dur}}(3\omega_{\nu}-1)
\end{equation}
In order to put observational constrain on parameter $\omega_{\nu0}$ and reconstruct the evolution of $\omega_{\nu}$, the above equation should be coupled to the autonomous equations of the three models. By doing this, we have reconstructed the evolution of $\omega_{\nu}$ using observational constraint. In Fig. 12, we depicted the evolution of  $\omega_{\nu}$ for three models. As can be seen,
neutrino equation of state parameter ($\omega_{\nu}(z)$) evolve from radiation dominated with value of $\omega_\nu=\frac{1}{3}$.
\begin{figure}
	\centering
	\includegraphics[scale=.45]{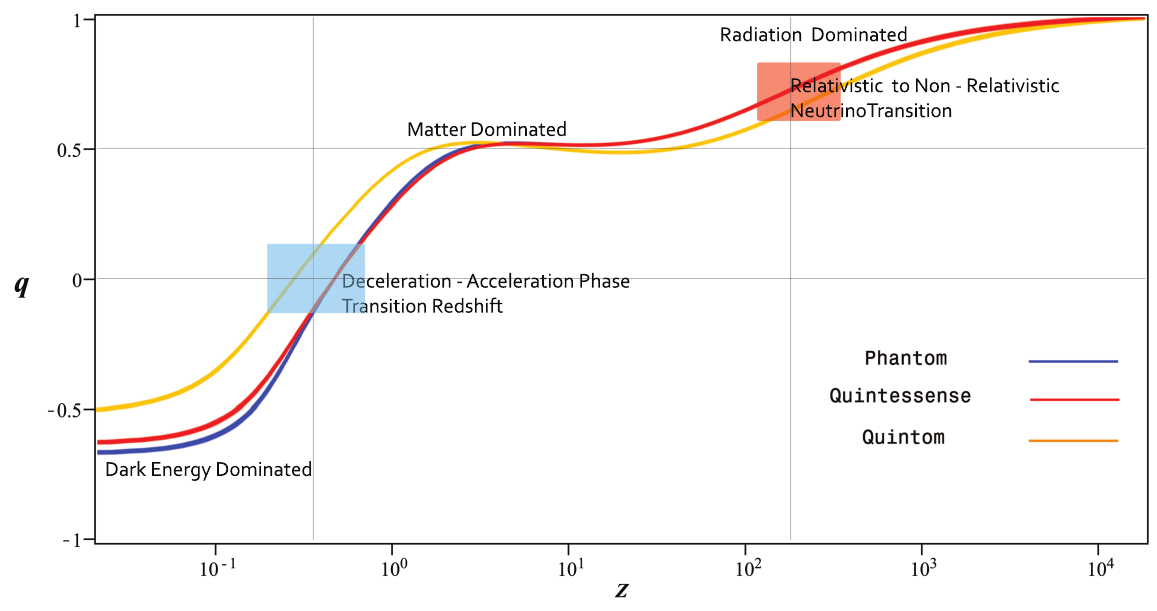}\hspace{0.1 cm}\\
	Figure 12: The plots show the evolution of of deceleration parameter q
	as a function of the redshift for the best-fitted values of the parameters in all three model
\end{figure}
\begin{figure}
	\centering
	\includegraphics[scale=.3]{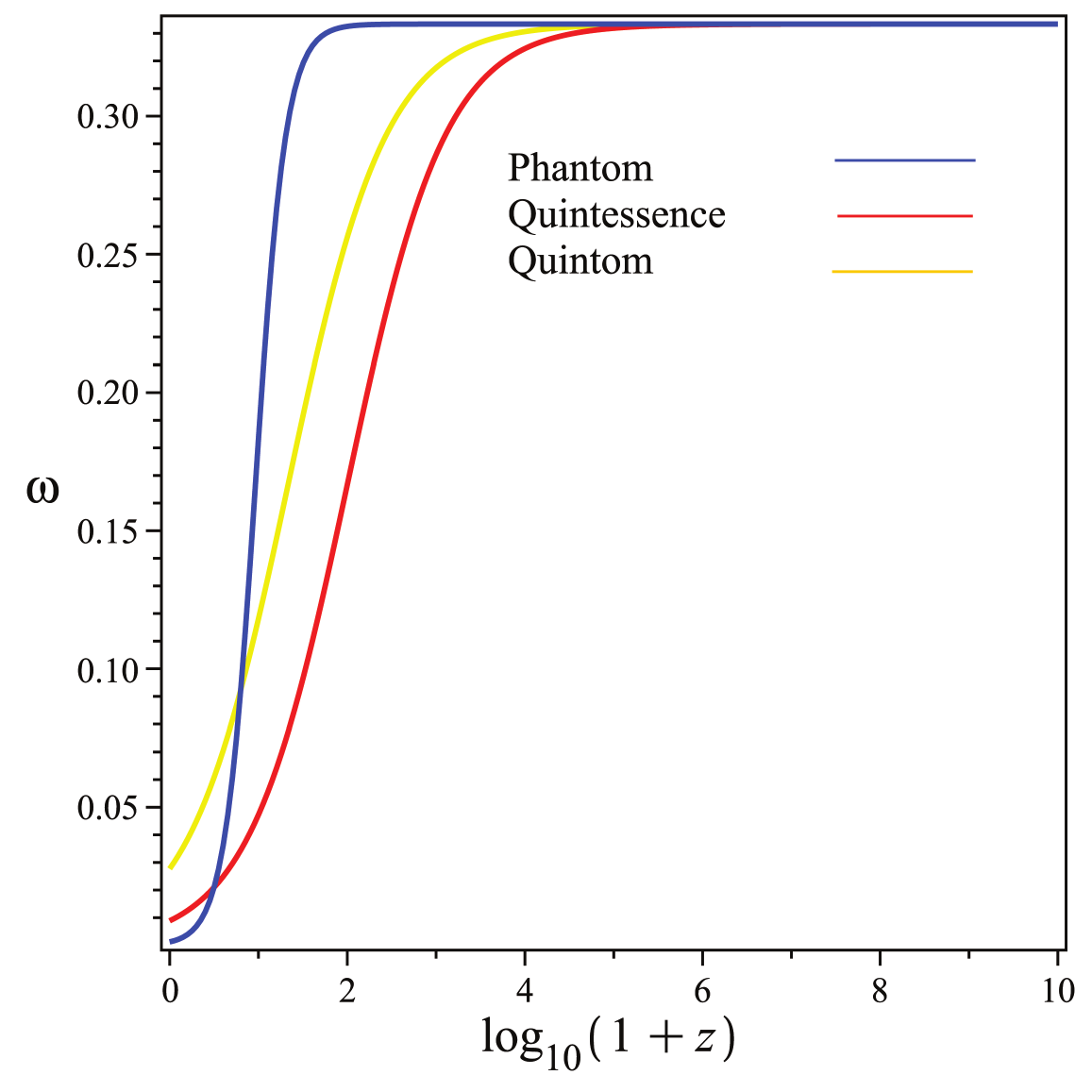}\hspace{0.1 cm}\\
	Figure 13: Figure shows the evolutions of equation of state parameters $\omega_{\nu} $ for all three models.
\end{figure}

Moreover, we desire to have a better control on the features of this
transition, namely, the epoch around which the transition is realized and the duration of realization.
We consider a semi-relativistic
phase (around $z_{\rm nr}$) between relativistic to non-relativistic transition, where the neutrino equation of state parameter ($\omega_{\nu}(z)$)
varies from 0 to $\frac{1}{3}$. To avoid rewriting the equations of each scalar field, we only show the results of our analysis in Figure 12. By doing this we obtained that during the
radiation dominated era $\omega_{\nu} = \frac{1}{3}$ and
in the recent universe $(\omega_{\nu} = 0.012, z_{\rm dur}= 3.61)$, $(\omega_{\nu} = 0.014,z_{\rm dur}= 3.63)$ and  $\omega_{\nu} = 0.026, z_{\rm dur}= 3.88$ for  quintessence model, Phantom and Quintom model respectively. The results are in good agreement with the results of \cite{Wali}.

\section{Conclusion}
The addition of massive neutrinos to the cosmological model alters both
the redshift evolution of $H(z)$, and the clustering of matter on small scales. By combining
different cosmological observables sensitive to these effects, it is forecasted that data from
upcoming surveys have the potential to reach the precision required to make the first
observational detection of neutrino mass \cite{Font-Ribera}.
In this paper, we used phantom, quintessence and quintom as dark energy and put a constraint on neutrino mass by coupling dark energy with neutrino.  We  first found that the total mass of neutrino is  $\sum m_{\nu}< 0.1197 eV$ $(95\% $ Confidence Level (C.L.)  for quintom model and $\sum m_{\nu}< 0.121 $eV $(95\% $ Confidence Level (C.L.) for phantom model and $\sum m_{\nu}< 0.122 $eV $(95\% $ Confidence Level (C.L.) for quintessence model. These results are in broad agreement with the results of Planck 2018 where the limit of the total neutrino mass is $\sum m_{\nu}<0.12 $eV ($95\%$ C.L., TT, TE, EE+lowE+lensing+BAO). Also, the interaction constant, $\beta$, for  these three models are investigated.\\
As we know, larger $\beta$ will generally lead to larger $m_{\nu}$ in the universe.\\
$\bullet$ In phantom model, the value of $\beta$  for combination data (Pantheon + CMB + BAO) is $ 0.229 $. This value indicate that the coupling between neutrino and dark energy is  small.\\
$\bullet$  In quintessence model,  $\beta$ is $ 0.63 $ which indicates that the dark energy neutrino interaction is grater than that of phantom model.\\
$\bullet$   $\beta$ value  in the quintum model is about $0.236$, which shows that in the world of quintom, the coupling  between neutrino and dark energy almost is the same as what phantom model has predicted.\\
What is more, the results obtained in this paper indicate that the value of the equation of state in the quintom model is $\omega_{\sigma}= -1.04 \ \ \  ,  \omega_{\phi}= -1$. By comparing the results obtained for the equation of state of these three models, it can be concluded that for the equation of state with a value of -1 or less (Quintom and Phantom models), the value of dark energy neutrino interaction is less than a situation when the  equation of state value is greater than -1 (Quintessence model). In what follows,  we surveyed $z_{\rm nr}$ for all three models and we obtained:\\

$\bullet$  For the phantom model(combination data)  we obtained  ${z_{\rm nr}} = 228.166$

$\bullet$  For the quintessence model(combination data)   we obtained  ${z_{\rm nr}} = 230.060$

$\bullet$  For the quintom model(combination data)  we obtained  ${z_{\rm nr}} = 225.704$\\

these results shows that the neutrinos become non - relativistic at distance almost $4282 $Mpc or $13.7$Gly Although this amount is much less than $z_{\rm eq}$.\\

Also, we indicated that the non - relativistic neutrino plays an important role in structure formation at the early universe. The results obtained in this paper for co-moving wavenumber (${k_{\rm nr}} \approx 0.018{\Omega _m}^{\frac{1}{2}}(\frac{{{m_\nu }}}{{1ev}})^{\frac{1}{2}}
h${Mp${c^{ - 1}}$})  are:\\
$\bullet$  For the phantom model(combination data) we obtained the non-relativistic neutrino wavenumber ${k_{\rm nr}}=0.0.000243$Mp${c^{-1}}$

$\bullet$  For the quintessence model(combination data) we obtained the non-relativistic neutrino wavenumber ${k_{\rm nr}}=0.000248$Mp${c^{-1}}$

$\bullet$  For the quintom model(combination data) we obtained the non-relativistic neutrino wavenumber ${k_{\rm nr}}=0.000245$Mp${c^{-1}}$.\\
These results are in general agreement with (\cite{Masatoshi};\cite{Christos}).\\
these results are in good agreement (\cite{Masatoshi};\cite{Christos}).\\
Furthermore, we consider the scenario when $\sum m_{\nu}$ equals to the minimum
value in each hierarchy and the lightest neutrino is massless. We work on the case that  all of the mass is in the third neutrino $(\theta = 1)$ or that
third neutrino is massless $(\theta = 0)$. In the case $(\theta = 1)$, we consider $\sum m_{\nu} = m_{3}$ and another flavors are zero. Results were obtained from analysis of this case is the same of results in  above scenario and is close to  inverted hierarchy (minimum mass of
0.1 eV). If we consider the case $(\theta = 0)$, $\sum m_{\nu}$ split to the two equal mass  and for each model we calculate the minimum mass which is very close to the minimum mass of the normal
hierarchy.
The results obtained for the normal
hierarchy are:\\

$\bullet$  For the phantom model(combination data)  we obtained  ${z_{\rm nr}} = 113.583$

$\bullet$  For the quintessence model(combination data)   we obtained  ${z_{\rm nr}} = 114.530$

$\bullet$  For the quintom model(combination data)  we obtained  ${z_{\rm nr}} = 112.352$\\
which is in the   matter dominate era. For  non-relativistic neutrino wave number\\
$\bullet$  For the phantom model(combination data) we obtained the non-relativistic neutrino wave number ${k_{\rm nr}}=0.000174$Mp${c^{-1}}$

$\bullet$  For the quintessence model(combination data) we obtained the non-relativistic neutrino wave number ${k_{\rm nr}}=0.000175$Mp${c^{-1}}$

$\bullet$  For the quintom model(combination data) we obtained the non-relativistic neutrino wave number ${k_{\rm nr}}=0.000173$Mp${c^{-1}}$\\

Also, we consider a semi-relativistic phase (around $z_{\rm nr}$) between relativistic to non-relativistic transition and we find that:\\
during the
radiation dominated era $\omega_{\nu} = \frac{1}{3}$  and
in the recent Universe $(\omega_{\nu} = 0.012, z_{\rm dur}= 3.61)$ for the quintessense model; $(\omega_{\nu} = 0.014,z_{\rm dur}= 3.63)$ for the phantom model ; $\omega_{\nu} = 0.026, z_{\rm dur}= 3.88$  for the quintom model which is consistent with results of the study carried out by   \cite{Wali}.

Moreover, we obtain the  $ \Delta $(AIC) Between 	$\Lambda$CDM and each model.\\
$\bullet$ The value of $ \Delta $(AIC) between the phantom model and $\Lambda$CDM  is 0.299\\

$\bullet$  The value of $ \Delta $(AIC) between the Quintessence model and $\Lambda$CDM  is 1.406\\

$\bullet$  The value of $ \Delta $(AIC) between the Quintom model and $\Lambda$CDM  is 6.021\\

As we can see in Table 4, the analyses based on the AIC indicated that there is more support for the interacting scalar fields with the neutrinos when compared to the $\Lambda$CDM  model, and ﬁt better than $\Lambda$CDM  for the full dataset combination, and improving the $\chi^2$.

\section{Data Availability}
The manuscript has no associated data or the data will not be deposited.

\end{document}